\documentstyle[12pt,aas2pp4, epsf]{article}

\def\Mdot{\hbox{$\dot {\rm M}$}}

\def\Lsun{\hbox{L$_\odot$}}

\def\Msun{\hbox{M$_\odot$}}

\def\kms{\hbox{km$\,$s$^{-1}$}}
\def\AV{\hbox{A$_{\rm V}$}}
\def\AJ{\hbox{A$_{\rm J}$}}
\def\AH{\hbox{A$_{\rm H}$}}
\def\AK{\hbox{A$_{\rm K}$}}
\def\AL{\hbox{A$_{\rm L}$}}
\def\BCK{\hbox{BC$_{\rm K}$}}
\def\BCV{\hbox{BC$_{\rm V}$}}

\makeatletter
\def\jnl@aj{AJ}
\ifx\revtex@jnl\jnl@aj\let\tablebreak=\nl\fi
\makeatother

\received{RECEIPT DATE}
\revised{REVISION DATE}
\accepted{ACCEPT DATE}
\journalid{VOL}{JOURNAL DATE}
\articleid{START PAGE}{END PAGE}
\paperid{MANUSCRIPT ID}
\cpright{TYPE}{YEAR}
\ccc{CODE}

\lefthead{Figer et al.}
\righthead{Massive Stars}
\begin{document}

\title{Massive Stars in the Quintuplet Cluster}

\author{Donald F. Figer\altaffilmark{1},
Ian S. McLean\altaffilmark{1}, Mark Morris\altaffilmark{1}}

\authoremail{figer@astro.ucla.edu}

\altaffiltext{1}{University of California, Los Angeles, Division of Astronomy, 
Department of Physics \& Astronomy, Los Angeles, CA 90095-1562}

\begin{abstract}

     We present near-infrared photometry and K-band spectra of newly-identified massive stars in the
Quintuplet Cluster, one of the three massive clusters projected within 50 pc of the Galactic Center. 
We find that the cluster contains a variety of massive stars, including more 
unambiguously identified Wolf-Rayet stars than any cluster in the Galaxy, and
over a dozen stars in earlier stages of evolution, i.e., LBV, Ofpe/WN9, and OB supergiants. 
One newly identified star is the second ``Luminous
Blue Variable'' in the cluster, after the ``Pistol Star.''
While we are unable to provide certain spectral classifications for the five enigmatic
Quintuplet-proper members, we tentatively propose that they are extremely dusty versions of the
WC stars found elsewhere in the cluster, and similar to the dozen or so known
examples in the Galaxy. Although the cluster parameters are uncertain because of photometric errors and
uncertainties in stellar models, i.e., extrapolating intial masses and estimating ionizing fluxes, 
we have the following conclusions.
Given the evolutionary stages of the identified stars, the cluster appears to be
about 4$\pm1$ Myr old, assuming coeval formation. The total mass in observed stars is $\sim$ 10$^3$ \Msun, and the
implied mass is $\sim$ 10$^4$ \Msun, assuming a lower
mass cutoff of 1 \Msun\ and a Salpeter initial mass function.
The implied mass density in stars is at least a few thousand \Msun\ pc$^{-3}$. 
The newly-identified stars increase the estimated ionizing flux from this
cluster by about an order of magnitude with
respect to earlier estimates, to 10$^{50.9}$ photons s$^{-1}$, or roughly 
what is required to ionize the nearby ``Sickle'' HII region (G0.18$-$0.04). 
The total luminosity from the massive cluster stars is $\approx$ 10$^{7.5}$ \Lsun,
enough to account for the heating of the nearby molecular cloud, M0.20$-$0.033. 
We propose a picture which integrates most of the major features in this part of the sky, excepting
the non-thermal filaments. 
We compare the cluster to other young massive clusters and globular clusters, finding that it is unique in stellar 
content and age, except, perhaps, for the young cluster in the central parsec of the Galaxy. In
addition, we find that the cluster is comparable to small ``super star clusters.''

\end{abstract}

\keywords{stars: supergiant --- Galaxy: center --- HII regions --- 
open clusters and associations --- ISM: individual (G0.15-0.05) --- ISM: individual (G0.18-0.04)}

\section{Introduction}

Three extraordinary bursts of star formation have produced young clusters
in the Galactic Center (GC) in the past 10 Myr\markcite{fmm96,krab,nag95,cote96,ser98} 
(Figer, Morris, \& McLean 1996; hereafter FMM96;
Krabbe et al.\ 1995; Nagata et al.\ 1995; Cotera et al.\ 1996; Serabyn, Shupe, \& Figer 1998). 
In addition to their location in the Galaxy, these clusters are also special 
for being the most massive young clusters in
our Galaxy. They offer us the possibility for investigating
massive star formation in 
molecular clouds with extra-solar metallicity, large internal turbulent velocities,
and strong magnetic fields. 
The many stars in each cluster may also finally provide the necessary statistics
to investigate the existence of a true upper mass cutoff to the initial mass function (IMF) in
the Galactic Center.

The Quintuplet Cluster is one of these massive clusters, and it is located approximately
30 pc, in projection, to the northeast of the Galactic Center\markcite{glass87} 
(Glass, Catchpole, \& Whitelock 1987). In addition to the five 
bright stars for which the Quintuplet was named\markcite{nag,oku} (Nagata et al.\ 1990;
Okuda et al.\ 1990), there is a clustering of hundreds of other stars in the
vicinity. Our JHK$^\prime$ color composite (Figure 1) clearly shows a concentration 
of bright stars spanning a diameter of $\approx$ 50$\arcsec$. Beyond
this, cluster members are indistinguishable in continuum images
from the already crowded field of stars in this part of the sky.

The cluster was originally observed as one or two sources in various 
infrared surveys\markcite{oku} (See \S1 in Okuda et al.\ 1990 for a summary), until 
the late eighties, when three groups finally
separated the cluster into a dozen or so sources\markcite{nag,oku,gla} 
(Nagata et al.\ 1990; Okuda et al.\ 1990; Glass, Moneti, \& Moorwood 1990). 
Nagata et al.\markcite{nag} (1990) and Okuda et al.\markcite{oku} (1990) noted
that 5 of the cluster stars are extraordinary in
their large luminosities, very cool spectral energy distributions, and lack
of intrinsic spectral features. They speculated that these objects might be
young, dust-enshrouded stars. 

Since then, many other cluster members have been identified as post main sequence
descendants of O-stars\markcite{mon,geb,har,FMM95,FMM96,cote96,f98a} 
(Moneti, Glass \& Moorwood 1994; hereafter, MGM94; Geballe et al.\ 1994; 
Harris et al.\ 1994; Figer, McLean \& Morris 1995; hereafter FMM95; FMM96; Cotera et al.\ 1996; Figer et al.\ 1998a). 
Most of these cluster stars are OB supergiants with strong winds, i.e, Wolf-Rayet stars, and LBVs. 
It is now clear that the Quintuplet Cluster is very massive, and that its members 
form a true physical group.

The Quintuplet Cluster is proving to be key in understanding answers to many important questions: 1)
Is the cluster in the inner parsec of the Galaxy really ``unique'', i.e., are
extraordinary physical processes or continuous star formation scenarios 
required to explain the stellar content of the
central cluster? 2) Do hot stars in the Quintuplet ionize the nearby \ion{H}{2} regions, G0.18$-$0.04 
(the ``Sickle'') and G0.15$-$0.05 (the ``Pistol'')? 3) Does the 
IMF in the Galactic Center favor the formation of high-mass stars? 
4) Are stars in Quintuplet consistent
with stellar evolution models which predict that WR/(WR+O) and WC/(WR+O) should be elevated
in higher metallicity regions?  

In this paper, we expand upon the work of FMM96 and Figer et al.\markcite{f98a} (1998a) 
in showing that the Quintuplet Cluster is extraordinary for its membership of 
massive evolved stars. We present new JHK$^{\prime}$nbL photometry (nbL is ``narrow-band L'') 
and K-band spectroscopy of the massive
stars in the Quintuplet with the goal of answering some of the aforementioned questions. In
\S 2, we describe the observations and data reduction. In \S 3, the spectral
classifications are presented.
We ``sum'' the properties of the individual stars in \S 4 to infer cluster properties,
i.e., mass, age, and ionizing flux; the latter is used to argue that Quintuplet stars are
ionizing the nearby HII regions. We compare the cluster properties to those of other
massive clusters in the Galaxy and Magellanic Clouds in \S 5, arguing that it is a ``near-twin'' to the young
cluster in the central parsec of the Galaxy (hereafter referred to as the ``Central Cluster''),
and is an older ``brother'' of the nearby Arches cluster. In this section, we present a consistent
picture to account for the proximity of the Quintuplet, Pistol, and Sickle.
We also interpret the Quintuplet Cluster as a ``super star cluster,'' i.e., those
seen around other galaxies. Finally, in \S 6, we present
conclusions, noting that the current sample of stars does not permit an 
investigation of the IMF or massive star evolutionary models. Such investigations 
await our HST/NICMOS data\markcite{f98c} (Figer et al.\ 1998c)
 
\section{Observations and Data Reduction}

\subsection{Observations}

     All data were taken with the UCLA double-beam near-infrared
camera\markcite{mcla,mclb} (McLean et al.\ 1993; McLean et al.\ 1994) at the University of
California Observatories' 3-m Shane telescope, 
producing a plate scale of 0\farcs7 pixel$^{-1}$. A grism was inserted into the beam, in conjunction
with a 2-pixel wide slit mask, to produce spectra covering the K-band atmospheric window and having  
R = $\lambda$/$\Delta\lambda_{\rm 2 \, pix}$ $\approx$ 525\markcite{figer} (Figer 1995). 
Table 1 gives the identifications and coordinates for the target stars, and the dates when the spectra were obtained.
The numbering results from a count of stars in the July 20, 1994,
K$^\prime$-band image, proceeding from East to West and South to North. Stars \#577 and \#578 are located
outside of this image, so they are numbered sequentially starting with the next integer after the
last star in the image. Figure 1 is a
JHK$^\prime$ color composite of an $\approx$ 2\farcm8 $\times$ 2\farcm8 region centered on
the cluster. Figures 2a-2d contain individual JHK$^\prime$nbL images. Figure 3a shows
an alternate greyscale stretch of the K$^\prime$ image shown in Figure 2c at an expanded scale. Figure 3b
gives a ``zoomed'' finder chart for the brightest cluster stars in Figure 3a which were included
in the slit-scan data cube (see below). 

\subsection{Photometry}

Photometry was extracted from two sets of images; the first set contains H- and K$^{\prime}$-band images 
obtained on July 20, 1994, and the second set contains JHK$^{\prime}$nbL images obtained on July 4, 1996.
The seeing generally produced images with less than 2 pixels (1\farcs4) full-width at half maximum.
The data were reduced according to the procedures in Figer\markcite{figer} (1995). In summary,
bias structure and dark current were removed from target frames by subtracting 
``bias+dark'' calibration frames; these images had the same 
exposure parameters as the target frames, except that an opaque mask
was inserted into the beam when they were taken. Variations in system efficiency, especially due to
pixel-to-pixel differences in quantum efficiency, were removed by dividing the target image
by a ``flat-field'' image; these images were constructed by forming the median of a stack of 
frames, each taken with the telescope in a slightly different position. This ``dithering''
serves to eliminate objects in the final flat-field frame.

Final photometry was extracted using \\
DAOPHOT, a point spread function-fitting routine in IRAF\footnote[1]
{IRAF is distributed by the National Optical Astronomy Observatories,
which are operated by the Association of Universities for Research
in Astronomy, Inc., under cooperative agreement with the National
Science Foundation.}. Photometric apertures with 4\farcs2 diameter were used, i.e., 3 times
the size of the seeing disk. The results are listed in Table 2. K 
magnitudes were converted from K$^{\prime}$ using the relation in\markcite{wai} Wainscoat \&
Cowie (1992). We observed atmospheric standards from the list in Elias et al.\markcite{elias82} (1982). 
The inferred zero-points from the July 4 images varied considerably as a function
of time and airmass, i.e., the night was not photometric. We adopted the
flux measurements in the first data set (Figer 1995), instead, for the H- and K-bands; the effective zero-points
were estimated by minimizing the difference between the average photometry in the two
data sets for the stars listed in Table 2. The J and nbL 
zero-points were extrapolated to high airmass from the observations of the standard
stars. We assess a photometric error of $\pm$0.2 magnitudes for all bands.

\subsection{Spectra}

The spectra cover most of the K-band and were reduced using procedures described in Figer, McLean, \& Najarro\markcite{fmn97} (1997). Some of the spectra were
extracted from a slit-scan data cube covering a 1$\arcmin$ (EW) $\times$ 2$\arcmin$ (NS) area, as
shown in Figure 4. The telescope was stepped by 1$\arcsec$ along the east-west direction between exposures.
Star \#211\markcite{mon} (Quintuplet star \#3 according to MGM94) was used as an atmospheric standard in all cases. 
The observations presented in this paper, and those of others\markcite{oku,mon,nks,f98b} 
(Okuda et al.\ 1990; MGM94; Nagata, Kobayashi, \& Sato 1994; Figer et al.\ 1998b), show that \#211
is spectroscopically featureless.
The divided spectra were multiplied by a blackbody function, with T = 630\, K, to correct for the intrinsic 
energy distribution of \#211. We have found that this temperature is appropriate for the apparent
K-band spectral energy distribution of \#211\markcite{f98b} (see Figure 1 in Figer et al.\ 1998b).
The spectra generally have S/N $\gtrsim$ 30, i.e., features having $\vert$EW$\vert$ $>$ 3 \AA\ should
be detectable at a level of a few times the noise in the continuum.

\section{Spectral Classifications}

	Our spectral classifications are based upon K-band spectra. The final classifications
are given in Table 3, along with the estimated luminosity and ionizing flux for each star.
The Wolf-Rayet (WR) stars were classified using the atlas of Figer, McLean, \& Najarro\markcite{fmn97} (1997;
R $\approx$ 525). 
Their montage of Galactic WR K-band spectra has been reproduced in Figure 5. The ``OBI'' stars were classified 
using the atlases of\markcite{han} Hanson \& Conti (1996) and\markcite{tamb} Tamblyn et al.\ (1996). 
     
\subsection{Wolf-Rayet Stars}

	Figure 5 shows that the emission lines for the Galactic WR stars 
tend to follow the expected trend of greater equivalent width for higher
ionization species with earlier subtype. Spectral classification based upon K-band spectra is more
effective in distinguishing different subtypes amongst the WN sequence
than the WC sequence. WR108 (the only bona-fide WN9 star in the sample) lacks \ion{He}{2} emission at 2.189 \micron,
while WN8 stars show a hint of it. This is the strongest feature in earlier
WN stars. This line, along with the 2.11 \micron\ feature
(\ion{He}{1}/\ion{C}{3}/\ion{N}{3}), allows accurate discrimination
to within 1 subtype for the WN sample in Figure 5 \\ 
(W$_{2.189\, \micron}$/W$_{2.11\, \micron}$).
The WC stars tend to have similar spectra except for the
latest types (WC8 and WC9). This degeneracy can be partially lifted by measuring flux
in the 3.09 \micron\ \ion{He}{2} line\markcite{figer} (Figer 1995). 
WR112 and WR118 have featureless spectra, presumably due to dilution by dust 
emission\markcite{wil} (Williams, van der Hucht \& Th\'e 1987). 

     The WR stars in our sample were classified by comparing their spectra (Figures 6a-6d) to the
spectra in Figure 5 and by comparing their flux excesses at 3.09 \micron\ to
those measured in Galactic WR stars. 
The newly identified WN9 stars in the Quintuplet Cluster, \#256
and \#274, have line widths similar to \#320 (FMM95-1), and they lack 
\ion{He}{2} emission (2.189 \micron). The new WN6 star, \#353e, has prominent
emission at 2.189 \micron\ which is comparable to the emission line strength
near the 2.166 \micron\ \ion{He}{2} line; it also has a considerable excess at 3.09 \micron. 

The new WC stars all have similar spectra, lacking the prominent emission at 2.058
\micron\ which is usually seen in WC9 stars, i.e., \#76\markcite{fmm95} (FMM95).  Two of the stars, \#309
and \#235, are classified as earlier than WC8 for their excesses at 3.09
\micron, while \#151 has very little excess there. Together, the
WN6 and the two ``$<$WC8'' stars are amongst the hottest identified
stars within 50 pc of the Galactic Center, although their ionizing
fluxes are quite meager owing to their small radii\markcite{crow1} (Crowther \& Smith 1996).

\subsection{OB Supergiants}

Spectra for the OB supergiants are shown in Figures 7-8. The ``$<$B0I'' stars have a featureless continuum, i.e.,
no features having $\vert$EW$\vert$ $>$ 2 \AA.     
Their K-band magnitudes put them in the supergiant class, and their
featureless spectra are consistent with a classifcation earlier than B0I. Spectra for later types, i.e., A- or F-
supergiants, have
a relatively strong Brackett-$\gamma$ line in absorption or emission\markcite{han} (Hanson et al.\ 1996).
All other OB supergiants were classified for the strengths of the three
primary lines in the K-band: Brackett-$\gamma$ (2.166 \micron),  \ion{He}{1} (2.058 \micron),
and \ion{He}{1} (2.112/2.113 \micron). The strength of the 2.058 \micron\ feature
is suspect when there appears to be features near 2.312 \micron\ and 2.370 \micron; all three features,
when present, may be due to incomplete atmospheric correction. The ``OBI'' stars
have Brackett-$\gamma$ and \ion{He}{1} (2.058 \micron) in
emission with \ion{He}{1} (2.112/2.113 \micron) in absorption; again, the 2.058 \micron\ line
might be contaminated by improper atmospheric correction, as described above. The spectra are
similar to those of HD 207329\markcite{tamb} (B1.5IB:e; Tamblyn et al.\ 1996) and BD+36 4063\markcite{tamb,han}
(ON9.7Ia; Tamblyn et al.\ 1996 and Hanson \& Conti 1996); it should
be noted, though, that ON9.7Iab stars in\markcite{han} Hanson \& Conti (1996) have
all three diagnostic features in absorption. Some Galactic Center stars also have similar
spectra\markcite{gen} (c.f.\ IRS16NE, IRS16NW, and IRS33E in Genzel et al.\ 1996). 
The early BI stars were classified by measuring the equivalent widths in these three spectral lines,
all in absorption, and comparing these values to those in the atlases.
The Ofpe/WN9 stars have been previously classified\markcite{geb,figer,cote96} (Geballe et al.\ 1994;
Figer 1995; Cotera et al.\ 1996).

\subsection{The Quintuplet-proper Members}

     The nature of the Quintuplet-proper members (QPMs) has remained a mystery
since their discovery. They are very bright in the infrared, m$_{\rm K}$ $\approx$ 6 to 9,
and have cool apparent spectral energy distributions, i.e., infrared color temperatures
between $\approx$ 600 to 1,000\, K. 
We present de-reddened SED's with blackbody fits in Figures 9a-d. The fits have been made
to the data, dereddened by \AK\ = 2.7, instead of the higher value (see
\S 4) derived for the other cluster stars; this has been arbitrarily done because the lower
extinction value is better fit by a single-temperature blackbody. There is no a priori 
reason to expect that the emission should follow a single-temperature blackbody. In
fact, the increased dereddened J-band flux that the stars would have if dereddened by
\AK\ = 3.2 might be an indication of a rising spectrum for the underlying photosphere.
We find good fits
for temperatures of 780\, K to 1,315\, K. Glass et al.\markcite{gla} (1990) performed a
similar analysis and found somewhat cooler temperatures due to the smaller
extinction value they assumed. 
After dereddening, their integrated infrared luminosities
are in the range 10$^{4.3}$ to 10$^{5.2}$ \Lsun, yet no spectral features 
characteristic of supergiants have been found. In fact, the objects are spectroscopically featureless
at all wavelengths observed, making their spectral classification ambiguous\markcite{nag,oku,gla} 
(Nagata et al.\ 1990; Okuda et al.\ 1990; Glass et al.\ 1990). 
Apparently, each object is composed of a powerful star(s) surrounded by dust. 

Some have suggested that these objects are protostars, or at least
not normal giants or supergiants\markcite{oku,nag,gla} (Okuda et al.\ 1990; Nagata et al.\ 1990; 
Glass et al.\ 1990); however, protostars would be much younger than the other
stars identified in the cluster. 
In addition, protostars often emit polarized light,
which is not the case for the QPMs\markcite{nks} (Nagata et al.\ 1994). 
Could they be OH/IR stars? Nagata et al.\markcite{nag} (1990)
make strong arguments against this hypothesis, the strongest being that the stars do not
have OH masers\markcite{hab,winn,lor} (Habing et al.\ 1983; Winnberg et al.\ 1985; Sjouwerman 1998).
In addition, OH/IR stars: 1) have deep water and CO absorption bands in their near-infrared spectra, 
2) are less luminous than the QPMs, 3) are warmer than the QPMs, and 4) are much older than the QPMs,
assuming that they are coeval with other cluster stars. Could they be red supergiant
``monsters,'' i.e., VY CMa? Such stars usually have SiO or water masers and deep
water absorption features in the near-infrared. Could they be embedded OB stars?
Again, this proposition has the same problem as the idea that they are protostars, i.e., 
such massive stars should have finished contracting and cleared away their circumstellar environment in much
less than 1 Myr\markcite{stahler} (Stahler 1994).

We suggest that these objects are 
dusty, late-type, WC stars\markcite{abb,wil,coh,fmm96} (``DWCL''; c.f. Abbott \& Conti 1987; Williams, van der Hucht \& 
Th\'e 1987; Cohen 1995; FMM96). DWCL stars represent a short-lived phase 
of evolution when the coolest WC types (WC8 and WC9) tend to form 
dust shells. Williams, van der Hucht \& Th\'e\markcite{wil} (1987) found that 19/27 of the
WC8 and WC9 stars they studied have significant circumstellar dust emission which can
be fit by blackbodies with temperatures of 780\, K to 1,650\, K (c.f., WR112 and WR118 
in Figure 5).

To test this hypothesis, we have calculated apparent K-band magnitudes that various Galactic WC9 stars 
would have if they were in the Quintuplet. Despite their common luminosities of
$\approx$ 10$^5$ \Lsun, we find that m$_{\rm K}$ would span a very large range, $\approx$ 3 (WR112) to 12 (WR92). This
is due to the different amounts of thermal emission from circumstellar dust around each
star. The QPMs have apparent magnitudes in this range, m$_{\rm K}$ $\approx$ 6$-$9.
Of course, the range for DWCLs is so large that this coincidence simply provides a consistency check, not a proof. 

As another test of our hypothesis, we obtained 
J-band spectra of the QPMs so that the classical emission-line spectra 
might be seen. All known DWCLs begin showing the expected WCL emission-line spectrum in
the J-band. We found no evidence for emission lines in the J-band spectra of these stars. Instead,
we observed that the flux is still decreasing strongly with decreasing wavelength, as might
be expected on the Wien side of a blackbody distribution. While this disagrees with the spectral
energy distributions of all known DWCLs, note that such stars have all been found in surveys detecting
their emission-line spectra. In other words,
perhaps there are DWCLs which are completely enshrouded, i.e., their emission-line spectrum
is not observable.

The nature of these stars is important for determining the WC/WN ratio in the cluster,
a number which provides a crucial test of stellar evolution models\markcite{mey} (c.f.\ Meynet 1995).
If they are DWCLs, then they are dustier than any others, begging the question: Is there
something special about the Galactic Center environment, such as its high metallicity, which causes the winds of 
DWCLs to be particularly dusty? If they are not DWCLs, then they represent a new
phenomenon. The same logic applies to the mid-infrared sources in the Central 
Cluster\markcite{bec} (Becklin et al.\ 1978).
One possible avenue for further investigation is to concentrate on what is
directly observable, i.e., the outer dust shell. We are pursuing this with high resolution
mid-infrared imaging to measure the sizes of the dust shells as a function of wavelength for
the QPMs and template DWCLs. 

\subsection{Luminous Blue Variables}

Luminous Blue Variables\markcite{cont84} (LBVs; Conti 1984) are rare stars in a presumably short phase of evolution
between the main sequence and the Wolf-Rayet (WR) phase\markcite{hd94}. They number about a half-dozen in the
Galaxy and an equal number in the Magellanic Clouds. Given the similarities of the Central
Cluster to the Quintuplet Cluster, it is perhaps no coincidence that both contain LBV or stars with
LBV-like spectra\markcite{tamb} 
(Tamblyn et al.\ 1996). FMM95 first identified the Pistol Star (\#134) as
an LBV candidate for its location in the HR diagram, near-infrared spectrum, and spatial proximity to the
Pistol Nebula. They also suggested that the star ejected the gas now seen in the surrounding Pistol Nebula. 
Figer et al.\markcite{f98b} (1998b) presented further near-infrared spectroscopy
and photometry and applied wind-atmosphere and stellar evolution models to argue that the star
is truly in an LBV stage. 

Another cluster star, \#362, appears to be luminous, hot, and photometrically variable. It brightened by
+0.75 in the K-band while becoming redder by +0.34 magnitudes in H$-$K 
between the time when the two data sets were taken. The coincidence of the star becoming
redder while also becoming brighter could be explained if it has a cool Mira-type companion.
It is also possible that the star was moving to the red in the HR diagram, as LBVs often do when entering
an eruptive stage. We favor the latter interpretation based upon our K-band spectrum (Figure 7a) which 
lacks CO absorption features expected from a Mira companion; the spectrum was obtained shortly after the
star had brightened, so it should be a fair representation of the spectral energy distribution
when the star was brighter. Note that such a large change in H$-$K is equivalent to a 
change in effective temperature from $\geq$ 30,000\, K to $\sim$ 3,000\, K for normal stars. 
Although this range in temperature is rather large, it could be explained by the
errors in the photometry. We assume that the fainter photometry
is a better representation of the flux emitted by a hot photosphere, so we use it to 
estimate the luminosity assuming \BCK\ = $-$1.5, the ``average'' for the LBVs discussed in
Blum, DePoy, \& Sellgren\markcite{blum95b} (1995b).

\subsection{Two nearby stars}

We obtained K-band spectra and photometry for two stars within 5$\arcmin$ (12.5 pc projected)
of the Quintuplet which we suspect of being very young (Figure 10). 
The stars are N21 and N42 in the list of Nagata et al.\markcite{nag93} (1993)
and were selected because they fall to the red side of the reddening vector in a color-color
plot, i.e., they are intrinsically red. We expected
that they might be similar to the Quintuplet proper members. Integrating their spectral energy
distributions gives L $\gtrsim$ 10$^4$ \Lsun.

The two stars have nearly featureless spectra,
except, perhaps, for emission at 2.06 \micron; this feature could be due to the \ion{He}{1} line
or incomplete cancellation of atmospheric absorption. We favor the first possibility, because
the latter usually manifests itself as a combination of absorption and emission features, similar in appearance
to P-Cygni profiles. N42 might have some weak emission features near 2.08 \micron\ (\ion{C}{4}?)
and 2.11 \micron\ (\ion{C}{3}). N21 appears to have absorption lines at 2.125 \micron, 2.182 \micron,
and 2.315 \micron. The latter is most likely due to imperfect atmospheric correction. We do not
have candidate transitions to associate with the other two lines. Both spectra are well-fit by
cool blackbodies (T $\approx$ 2,000 to 2,500 K) after applying dereddening to account for \AK\ = 3.2; however
the extinction is not known. A larger assumed extinction would tend to produce a hotter energy
distribution, but the dereddened spectra would then show an excessively high flux at shorter 
wavelengths. This is the expected effect in trying to deredden an intrinsically cool spectrum
by assuming too much extinction. 

The energy distributions and emission lines favor 
the possibility that these are hot stars embedded in dust. 
It is unclear if these stars were formed in the same or related star formation events.
They might be outlying members of the Quintuplet Cluster, but it is difficult to determine without
having velocities or ages for the stars. 

\section{Cluster Properties}

	In this section, we sum the individual contributions to the cluster properties
in order to allow a comparison to other massive clusters and to determine whether the cluster is heating and ionizing
nearby clouds and HII regions. For most parameters, we give the observed quantities as well
as the implied values, assuming a Salpeter IMF\markcite{salpeter} (Salpeter 1955), and
a particular lower mass cutoff as discussed below. Most of the estimates depend upon luminosities of the 
individual stars, so we start by estimating the extinction and distance.

\subsection{Extinction}

Figer et al.\markcite{f98b} (1998b) reviewed extinction estimates to the cluster for their
study of the Pistol Star, estimating \AK\ = 3.28. The estimate represents an average value
inferred from color excesses of the stars in Table 1. Table 4 tabulates apparent colors and 
color excesses for the hot stars identified in the cluster. We consider two sample groups, the ``B1I$-$B3I'' 
stars (N = 5), for which the spectral classifications are the most precise in the sample, and 
the ``OBI'' stars (N = 9); the latter includes the former.
Although the ``OBI'' stars potentially span a larger range in subtype, i.e.\
O3$-$B9, they span a very narrow range in colors\markcite{koor83} (c.f.\ Koornneef 1983), so we assume
that all such stars have similar colors. 

We use \AK\ = 3.28$\pm$0.5 (\AV\ = 29$\pm$5) \\ 
throughout this paper, unless otherwise noted,
where the error is the quadrature sum of the standard deviation of (H$-$K) for the full sample of OB supergiant stars in the cluster and the photometric errors. The variation in (H$-$K) for these stars is due to differences in the
apparent colors of the stars, i.e., it is not due to inaccurate photometry or confusion. This
can be seen in K-band spectra which show that \AK\ = 3.28 would overestimate the redenning
to some of the stars, i.e., their dereddened energy distributions would be greater than
that of an infinite temperature blackbody. We take this to indicate that there is some
differential extinction across the field, and that an extinction value for each star
will eventually have to be individually computed; however, the Quintuplet proper members are
probably intrinsically very red. (We cannot estimate the true error in our estimate 
because it is dominated by systematic effects in the extinction law.)

\subsection{Distance}

We argue that the cluster is at the distance of the Galactic Center\markcite{Reid93} 
(d$_{\rm GC}$ = 8,000 pc; Reid 1993) for four reasons. 

First, V$_{\rm LOS}$ $\approx$ 130 \kms\ for the cluster stars
in every case where it has been measured\markcite{figer} (Figer 1995);
this is also true for gas in the Pistol Nebula\markcite{yus,figer,lang97} (Yusef-Zadeh, Morris, \& van Gorkom 1989;
Figer 1995; Lang et al.\ 1997). Such a high velocity is unlikely
for objects along the line of sight to the Galactic Center, and a value of this magnitude is 
expected for an object with an orbital radius equal to the projected radius
of the Quintuplet from the Galactic Center, i.e., v$_{\rm orbital}$ = (GM/R)$^{1/2}$ $\approx$ 150 \kms,
where M = 10$^{8.2}$ \Msun\ is the enclosed mass inside a 30 pc 
orbit\markcite{sell} (Sellgren et al.\ 1989).

Second, the Quintuplet Cluster appears to be ionizing the nearby
``25 \kms\ cloud''\markcite{lly89,sg91} (M0.20$-$0.033; Lasenby, Lasenby, \& Yusef-Zadeh 1989;
Serabyn \& G\"usten 1991), thus creating the Sickle\markcite{yzm,figer96,lang97} 
(Yusef-Zadeh \& Morris 1987; Figer 1996; Lang et al.\ 1997). 
The particularly large line-widths
associated with the molecular cloud are consistent with a location within the central molecular zone of 
the GC\markcite{sm94} (Serabyn \& Morris 1994).

Third, the inferred extinction (see above), interstellar polarization, and silicate absorption produced
by the intervening interstellar medium to the cluster are consistent with a location at
the GC\markcite{oku} (Okuda et al.\ 1990).

Lastly, components of the CO absorption bandhead at 4.66 \micron\ in the QPMs' spectra have been attributed to
the 250 pc ring\markcite{okud} (Okuda et al.\ 1990), due to their velocity shifts.

\subsection{Luminosity}

The total cluster luminosity can be estimated by summing the individual luminosities
of the identified stars. This will provide a lower limit, because the
cluster probably contains many unseen lower mass members; however, the total 
cluster luminosity should be dominated by massive stars for a reasonable IMF. 
We assumed a distance, extinction, and bolometric correction for each star, and
tabulated the results in Table 3. It is important to note that this estimate assumes
that the cluster is coeval, and that the present-day masses are well-determined by the
spectroscopic analysis above.

In order to estimate luminosities, we estimate the bolometric correction at K, \BCK\ = M$_{\rm BOL}$ $-$ M$_{\rm K}$. 
For the Ofpe/WN9 stars, we assume \BCK\ = $-$2.9\markcite{naja97} (Najarro et al.\ 1997).
We assume \BCK\ = $-$3.3 for the WC stars and \BCK\ = $-$2.9 for the WN stars\markcite{crow1} (Crowther \& Smith 1996). 
We integrate the infrared flux for the DWCL stars. 
For the OB supergiant stars, we assume \BCK\ = $-$2.0. This is somewhat conservative
compared to \BCK\ = $-$2.4, which is found by subtracting V$-$K\markcite{koor83} (Koornneef 1983) from 
\BCV\ for O9I stars\markcite{vac} (Vacca, Garmany, \& Shull 1996); however, it
allows for the fact that some of the OB supergiant stars will be later than late O type. For the LBVs, we
assume \BCK\ = $-$1.5, according to the analysis of the Pistol Star by Figer et al.\markcite{f98b} (1998b).

The total cluster luminosity is 10$^{7.5}$ \Lsun, counting all the stars in Table 3, and using
the lower luminosity limit for the Pistol Star. Two of the most luminous stars in the cluster are LBVs, stars which
span a very large range in \BCK; but even without these two stars, we
find L$_{\rm cluster}$ = 10$^{7.3}$ \Lsun. Either value suggests that the Quintuplet Cluster is
responsible for heating M0.20$-$0.033, which emits
10$^7$ \Lsun\ of radiation at infrared wavelengths\markcite{morris} (Morris, Davidson, \& Werner 1995).

\subsection{Mass}

	The total cluster mass cannot be directly estimated because the mass function has
not been measured. The following analysis makes the assumption that the cluster has an 
IMF slope which is nearly Salpeter\markcite{salpeter} (Salpeter 1955). Note that measured
IMF slopes vary considerably about the Salpeter value\markcite{scalo98} (Scalo 1998).

     We estimate the total cluster mass by counting the number of stars with initial masses
between an assumed upper mass cutoff of 120 \Msun\ and the lowest initial mass inferred from the list of
stars identified in Table 1, $\approx$ 20 \Msun. Assuming 30 stars in this mass range, 
we calculate M$_{\rm cluster}$
$\approx$ 10$^{4.2}$ \Msun\ for m$_{\rm lower}$ = 0.1 \Msun\ and 10$^{3.8}$ \Msun\ 
for m$_{\rm lower}$ = 1 \Msun.

We can compare this value to the mass required to ensure that the cluster
is bound against tidal disruption. Assuming a circular orbit with a velocity equal
to the line-of-sight velocity\markcite{figer} (130 km s$^{-1}$; Figer 1995), and an orbital
radius equal to the projected distance from the GC, we find an orbital time
of $\approx$ 1.5 Myr. The enclosed
mass at this radius is $\approx$ 10$^{8.2}$ \Msun\markcite{sell} (Sellgren et al.\ 1989). 
The condition for a star to be tidally bound at radius r$_{\rm cluster}$ is 
M$_{\rm total}$ $\gtrsim$ 2 $\times$ M$_{\rm r<30 \, pc}$ $\times$ 
(r$_{\rm cluster}$/30 pc)$^{3}$ = 10$^{4.1}$ \Msun, where r$_{\rm cluster} $
is the average distance of the stars in the table from the center of the cluster
and is $\approx$ 1 pc. This value is on the same order as the values above, suggesting
that the cluster is marginally bound against tidal disruption, at best.

\subsection{Age}

To estimate a single cluster age, we must assume that the cluster members in Table 1 are coeval.
The notion of coevality is ambiguous when contraction timescales widely range, such as is the 
case when comparing high mass to low mass stars. We are not
subject to this problem for the current analysis because we are concerned only with the massive
stars, which all have very short contraction times\markcite{stahler} (Stahler 1994). 

Model isochrones from the Geneva models\markcite{meyn94} (Meynet et al.\
1994) are shown in Figure 11 for twice solar metallicity and the ``2$\times$'' \Mdot\ models.
Data for the B supergiants
in Table 1 have been overplotted. The stars have been given the same temperatures in light of
their similar classifications, although note that the uncertainty in the spectral classifications
is not represented in the error bar.  Using lower metallicity 
or lower mass-loss rates will tend to give a higher age for the cluster, where the extreme value
of 6 Myr is given by solar metallicity and the ``standard'' \Mdot\ models.

The assumption of coevality is roughly consistent with the ages we estimate for the individual stars in
Table 1\markcite{mey} (Meynet 1995; and references therein).
The presence of WC stars requires that the cluster is older than 2.5 Myr. The sole
red supergiant requires an age $\gtrsim$ 4 Myr (assuming M$_{\rm initial}$ = 40 \Msun, 
``2$\times$'' \Mdot\ models, and Z=0.04).
The Pistol Star requires an age of $\lesssim$ 2.1 Myr, according to Figer et al.\markcite{f98b} (1998b).
The presence of OI stars requires an age between 2.5 and 4.7 Myr, 
depending on the mass-loss rates and the metallicity. We adopt an ``average'' age of 4 Myr for
the cluster.

\subsection{Ionizing Flux}

     Harris et al.\ \markcite{har}(1994) estimate that the Sickle requires a Lyman
continuum flux of Q$_{\rm Sickle} \approx$ 10$^{50.5}$ s$^{-1}$, and, according to 
Yusef-Zadeh et al.\markcite{yus} (1989), the Pistol requires 
Q$_{\rm Pistol} \approx$ 10$^{48.6}$ s$^{-1}$. Timmermann et al.\ \markcite{tim}(1996) use 
the radio flux at 32 GHz to estimate that the Quintuplet produces 
Q$_{\rm Quin, \ radio} \approx$ 10$^{50.2}/\eta$ s$^{-1}$, where $\eta$ is less than 1 and 
accounts for dust absorption and deviations from an ionization-bounded region. 

We find Q$_{\rm Quin, \ stars} \approx$ 10$^{50.9}$ s$^{-1}$ by summing the contributions from
individual stars in Table 3. The OB stars presented here 
represent some of the previously predicted population of ``unseen'' hot stars\markcite{har} (Harris 
et al.\ 1994; FMM95; Timmermann et al.\ 1996). While there may still be 
other O-stars with less pronounced spectral features in the cluster, i.e.\ those still on the main sequence, it appears that the presently identified hot stars produce enough ionizing radiation
to ionize the gas in the Sickle, even thought the geometry of the Sickle in relation to the
cluster implies a covering fraction of $\approx$ 1/3. 

Figer et al.\ \markcite{f98b}(1998b) give new estimates of the ionizing
flux emitted by the Pistol Star spanning
eight orders of magnitude (see their ``L'' and ``H'' models), demonstrating that 
it is highly dependent upon the exact physical parameters for the star. The range includes
the ionizing flux required to ionize the Pistol Nebula, although
contributions from other hot stars in the Quintuplet may be
important, i.e., \#151 (WC8) and \#235 ($<$WC8). 

\section{The Quintuplet Cluster in Context}

\subsection{The Quintuplet Cluster compared to other massive clusters}

Table 5 compares the Quintuplet Cluster and other massive clusters in mass, size, density,
age, luminosity, and Lyman continuum flux. The total mass of observed stars, ``M1,'' is subject to
observational effects, because of the varying distances of the clusters and the extent to
which they have been studied. ``M2'' is supposed to represent the total inferred mass of 
the cluster; note that our estimates are for an IMF truncated below 1 \Msun. The radius is
normally taken as the half-light radius, or average distance from the centroid of the
cluster. Because of the slightly different definitions used in the estimates, the values
should be taken as a rough indication of compactness; the same caution is obviously
applicable to the resultant densities, ``$\rho$1'' and ``$\rho$2,'' which are simply the 
mass estimates divided by the volume.

The Quintuplet Cluster is most comparable to NGC 3603 in mass and luminosity, but notice that
it is much less dense, probably older, and produces less ionizing flux. Perhaps the differences in
density and ionizing flux result from the difference in age, that is, the inevitable expansion of an
unbound cluster, and the natural decrease of ionizing flux with age. 
The former effect might
be amplified in the strong tidal field of the Galactic Center. Age might also be the reason
for the large number of WR stars in the Quintuplet Cluster (8 to 13) compared to NGC 3603 (3, of
the H-rich variety). In fact, the WR stars in NGC 3603 are probably hydrogen core burning
O-stars with particularly thick winds, like those found in R136\markcite{mass} (Massey \& Hunter 1998); such
stars are younger than ``true,'' i.e., He-burning, WR stars.

\subsection{The Quintuplet Cluster and the Central Cluster}

The emission-line stars in the central parsec of the Galaxy have been regarded as
``exotic'' for their spectral characteristics in the 
K-band\markcite{all,kraa,lib95,krab,blum95a,tamb} (Allen, Hyland
\& Hillier 1990; Krabbe et al.\ 1991; Libonate et al.\ 1995; 
Krabbe et al.\ 1995; Blum et al.\ 1995; Tamblyn et al.\ 1996). While similar stars can be found in the Galaxy
and the Large Magellanic cloud, they are rare and the ensemble of stars at 
the center, as a collection, is remarkable because it contains a concentration of such
striking stars.

The Quintuplet Cluster contains similar stars, as can be seen in Figure 4 and
FMM95. The WC9 stars found in the center\markcite{blum95b,krab} (Blum, Sellgren \&
DePoy 1995; Krabbe et al.\ 1995) have counterparts in the Quintuplet Cluster
(here and FMM95-2); in fact, IRS6 in the Galactic Center is a particularly dusty WC9. 
IRS16NE, which has LBV-like spectral characteristics\markcite{tamb}
(Tamblyn et al.\ 1996), is similar to the Pistol Star, and \#362. 
Some of the Ofpe/WN9 types in the center (IRS16 components) are similar to q8,
q10\markcite{geb,figer,cote96} (Geballe et al.\ 1994; Figer 1995; Cotera et al.\ 1996), 
and FMM95-1. IRS33E, IRS16NE, and IRS16NW have spectra
similar to the OB supergiant stars in the Quintuplet\markcite{figer,naja,gen} (Figer 1995; Najarro 1995;
Genzel et al.\ 1996). IRS7, the red supergiant in the central cluster, is similar 
to q7 in the Quintuplet\markcite{mon} (Moneti, Glass \& Moorwood 1994).
Finally, the QPMs share similar spectral energy distributions\markcite{oku,bec} (Okuda et 
al.\ 1990; Becklin et al.\ 1978) and K-band spectra\markcite{figer} (Figer 1995) with the very red sources in the
Galactic Center, c.f.\ IRS8; both groups of stars may be dusty, late-type, WC stars (DWCLs).

In total, both clusters have approximately 27 supergiants, and contain roughly equal numbers
of massive subtypes; therefore, the total mass in massive identified stars is the same in
both clusters, $\approx$ 10$^3$ \Msun (see Table 5). The small difference in total mass extrapolated
down to m$_{\rm lower}$ is due to the different techniques used to make this estimate\markcite{krab} (see
notes in Table 5 and Krabbe et al.\ 1995). The difference in ``$\rho$2'' is due to the difference
in radius for the two clusters, but this is likely owed to a selection effect. The central cluster has
been probed with very high resolution measurements, something which has not been done for the
Quintuplet Cluster. Indeed, there appears to be a dense clustering of bright stars in the 
north-south ridge situated between the four westernmost QPMs and \#211. Some of those stars have
already been identified as being very massive, but more identifications are limited by the
spatial resolution of current observations. Further identifications in this stellar concentration would lower the
effective radius for massive stars in the cluster. 
The collective luminosity and Lyman continuum flux for stars in each cluster are similar,
where differences might be attributable to modeling. 

In summary, the Quintuplet and Central cluster are nearly identical in mass and age. The primary
difference seems to be the enigmatic broad helium-line stars in the Central Cluster\markcite{kraa} (Krabbe et
al.\ 1991). However, those
stars are still evolved massive stars like the ones found in the Quintuplet Cluster, and the
detailed line profiles in the Central Cluster stars are likely due to a small difference in age between
stars in the two clusters\markcite{naja97} (Najarro et al.\ 1997).

\subsection{The Quintuplet, Pistol, Sickle complex}

The center of the Quintuplet Cluster is situated $\approx$ 10$\arcsec$ due north 
of G0.15$-$0.05 (the Pistol), and a few arcminutes to the southeast of G0.18$-$0.04 
(the Sickle)\markcite{yzm} (see Yusef-Zadeh \& Morris 1987 for nomenclature). 
Serabyn \& G\"usten\markcite{sg91} (1991) argued that the Sickle \ion{H}{2} region is the 
ionized surface of M0.20$-$0.033 (the ``25 \kms\ molecular cloud'').
Serabyn \& Morris\markcite{sm94} (1994) suggested that free electrons in the Sickle are accelerated by MHD
interactions at this interface, and resultant relativistic electrons
stream along nearby magnetic field lines, producing the non-thermal radio arcs. 
FMM95 suggested that the Pistol Nebula was ejected from, and is now
partially ionized by, the Pistol Star.

The data in this paper are consistent with the hypothesis that the Quintpulet Cluster
is heating dust in M0.20$-$0.033 and ionizing the gas in the Sickle. 
M0.20$-$0.033 emits $\approx$ 10$^7$ \Lsun\ at far-infrared wavelengths\markcite{morris} 
(Morris et al.\ 1995), consistent with the
idea that the Quintuplet Cluster and the molecular cloud are physically connected and not
just coincident along the line of sight. Note that the large difference in velocities for the 
cluster and the cloud indicates that the cloud probably did not spawn the cluster.

Lang et al.\markcite{lang97} (1997) presented radio recombination line 
data of both the Pistol and the Sickle showing that the
former might be enhanced in helium, while the latter has solar helium abundance. Again, this is
consistent with the idea that the molecular cloud is a wayward interloper in the region while
the Pistol Nebula is composed of ejecta which has been subject to stellar nucleosynthesis. They
also showed that ionized gas at the edges of the Sickle has been accelerated to higher velocities than
that near the center of the nebula and the molecular gas in the cloud. We interpret this
velocity field as indicating an interaction between the strong winds of the Quintuplet stars and
the surface of M0.20$-$0.033. This agrees with Simpson et al.\markcite{simp97} (1997) who 
suggest that the higher velocities come from a ``champagne flow'' 
interaction\markcite{lang97} (also see Lang et al.\ 1997)

A nearby ring of emission at \\
{\it l,b} = $(+0.15, -0.14)$
has been detected in mid-infrared data by the Midcourse Space Experiment\markcite{msx} 
(MSX; Shipman, Egan, \& Price 1996). 
This ring appears to be associated with the complex, being coincident on its northwest portion with 
the Sickle\markcite{egan} (Egan et al.\ 1998). 
The cluster is offset from the center of the bubble, but such an offset might be explained
by a relative motion of the Quintuplet Cluster and the local ISM, or to a large-scale density gradient in 
the medium into which the bubble is expanding.
Assuming that winds and radiation pressure from O-stars in the cluster have plowed the local
ISM into a dust shell, we might expect a bubble expansion timescale of 10$^4$ yrs (V$_{\rm exp}$ = 1,000 \kms).
Such a timescale is more reasonable for a shell which was swept up by an expanding supernova shock front.
Indeed, we expect 1 supernova per 50,000 yrs in the cluster, starting at a cluster age of $\approx$ 4 Myr
according to Meynet\markcite{mey} (1995). A further analysis of this dust ring 
is under way using ISO data to investigate the excitation and kinematics\markcite{lev} 
(Levine et al.\ 1998). 

\subsection{The Quintuplet as a ``Super Star Cluster''}

Ho \& Filippenko\markcite{ho} (1996; and references therein) argued that we are seeing present-day
globular cluster formation in other galaxies. These super star clusters are recognized for
their large luminosities, large masses, and compactness. Some have claimed that such
clusters are produced in turbulent environments. Table 5
includes the physical characteristics of globular and super star clusters for comparison with 
the three Galactic Center clusters. The
Quintuplet Cluster, Arches Cluster, and the Central Cluster are similar to the other 
objects compared in the table, except for their smaller mass, suggesting that they represent 
the low-mass end of the distribution of such objects.

If correct, then the GC clusters would be the closest examples of super star clusters. It would
also add evidence that such clusters can form in 
galaxies which are not interacting. This might suggest a unified model for such phenomena. 
Perhaps super star clusters are formed wherever dense molecular clouds collide, or are strongly shocked, 
to form dense, massive clusters. Such environments might be found in the early Galaxy and colliding galaxies, 
but they might also be found in the present-day Galactic Center. We speculate that the strength of the shock that
provokes the cluster formation is the primary determinant of the cluster mass. In that case,
the more violent encounters, such as those which occur in colliding galaxies, are the ones which
would naturally produce the most massive clusters.

\section{Conclusions}

Although many of the parameters derived in this paper rely upon only near-infrared data, we
conclude the following.
The Quintuplet Cluster is extraordinary for its content of massive
stars, containing more bona-fide Wolf Rayet stars than any other cluster in the Galaxy. 
It is the source of ionizing photons for the nearby \ion{H}{2} regions, the ``Pistol'' and the ``Sickle.'' 
The types of stars found in the cluster are consistent with a coeval population, and the
cluster age is $\approx$ 4$\pm1$ Myr, according to stellar evolution models. The total mass
in observed stars is $\sim$ 10$^3$ \Msun, and the inferred mass is $\sim$ 10$^4$ \Msun\ for
a Salpeter IMF and a lower mass cutoff in the range of 1 \Msun\ to 0.1 \Msun. The observed
mass density in stars is at least a few hundred \Msun\ pc$^{-3}$, and the inferred density is at least a few
thousand \Msun\ pc$^{-3}$.
The cluster is one of three massive clusters
in the Galactic Center which are similar in many respects to super star
clusters found in other galaxies. The current sample is limited to massive stars in various
stages of post main sequence evolution and their initial masses are difficult to estimate. 
An accurate determination of the IMF in this cluster must await deeper observations\markcite{f98c} 
(NICMOS/HST; Figer et al.\ 1998c) which will reveal main sequence stars, where the association between infrared flux and
initial mass is better known.

\acknowledgements

We thank the members of the Infrared Imaging Detector Laboratory at UCLA. We also
thank Eric E. Becklin for useful discussions concerning the extinction to the Quintuplet Cluster. 
We thank Paul Crowther for stimulating discussions concerning \BCK\ and the ionizing fluxes
of WR stars. We are very grateful to the referee, who provided many construvive comments which 
considerably improved the final manuscript.
Finally, we thank Paco Najarro for pioneering the efforts to model massive stars
in the Galactic Center.

\tiny
\clearpage
\begin{deluxetable}{rrrrrrl}
\tablecolumns{7}
\tablewidth{400pt}
\tablecaption{Log of Spectroscopic Observations}
\tablehead{\colhead{ID\#} &
\colhead{Nag90\tablenotemark{a}}    &  \colhead{MGM94\tablenotemark{b}}   & \colhead{R.A.\tablenotemark{c}} &
\multicolumn{2}{c}{Dec.\tablenotemark{d}} & \colhead{Dates Observed} \\
\cline{5-6} \\
\colhead{} & \colhead{} & \colhead{} & \colhead{$s$} & \colhead{$\arcmin$} & 
\colhead{$\arcsec$} }
\startdata
76 & \nodata & \nodata & 5.1 & 49 & 11.6 & July 26, 1994 \nl			
134\tablenotemark{e} & \nodata & seren & 4.8 & 48 & 56.9 & July 26, 1994 \nl			
151 & \nodata & \nodata & 4.4 & 48 & 53.8 & August 5, 1995 \nl			
157 & \nodata & \nodata & 3.5 & 48 & 52.0 & August 5, 1995 \nl			
178 & \nodata & \nodata & 9.3 & 48 & 46.2 & August 5, 1995 \nl			
192 & \nodata & 7 & 6.2 & 48 & 43.6 & \nodata \nl			
197 & \nodata & \nodata & 4.9 & 48 & 42.9 & August 5, 1995 \nl			
211\tablenotemark{f} & GCS4 & 3 & 5.5 & 48 & 39.0 & \nodata \nl			
231\tablenotemark{g} & GCS3-2 & 2 & 4.3 & 48 & 34.0 & \nodata \nl			
235 & G & 11a & 4.8 & 48 & 33.5 & August 5, 1995 \nl			
240 & \nodata & 8 & 5.5 & 48 & 31.2 & \nodata \nl			
241 & F & 10 & 4.7 & 48 & 30.2 & August 5, 1995 \nl			
242 & \nodata & \nodata & 4.1 & 48 & 30.0 & August 5, 1995 \nl			
243 & GCS3-4 & 1 & 3.7 & 48 & 29.9 & \nodata \nl			
250 & B & 6 & 5.0 & 48 & 27.9 & August 5, 1995, August 20, 1996 \nl			
251 & GCS3-1 & 4 & 4.4 & 48 & 27.6 & \nodata \nl			
252 & \nodata & \nodata & 4.1 & 48 & 27.1 & August 5, 1995 \nl			
256 & \nodata & \nodata & 6.1 & 48 & 25.2 & August 5, 1995 \nl			
257 & D & 13 & 4.8 & 48 & 25.5 & August 5, 1995 \nl			
258 & GCS3-3 & 9 & 3.9 & 48 & 24.7 & \nodata \nl			
269 & A & \nodata & 5.1 & 48 & 23.1 & August 5, 1995 \nl			
270N & C & 5 & 4.7 & 48 & 21.3 &  August 5, 1995 \nl			
270S & C & 15 & 4.7 & 48 & 21.9 & \nodata \nl			
274\tablenotemark{h} & \nodata & \nodata & 7.1 & 48 & 22.4 & August 5, 1995 \nl			
276 & \nodata & \nodata & 3.0 & 48 & 22.3 & August 5, 1995 \nl			
278 & E & 12 & 4.7 & 48 & 27.5 & August 5, 1995 \nl			
301 & \nodata & \nodata & 5.6 & 48 & 15.0 & August 5, 1995, August 20, 1996 \nl			
307 & \nodata & \nodata & 5.1 & 48 & 13.5 & August 5, 1995, August 20, 1996 \nl			
309 & \nodata & \nodata & 7.1 & 48 & 12.1 & August 5, 1995 \nl			
311 & \nodata & \nodata & 3.3 & 48 & 12.5 & August 5, 1995, August 20, 1996 \nl			
320 & \nodata & \nodata & 3.7 & 48 & 9.7 & July 26, 1994 \nl			
344 & \nodata & \nodata & 6.3 & 48 & 2.6 & August 5, 1995, August 20, 1996 \nl			
353E & \nodata & \nodata & 0.8 & 47 & 58.3 & August 3, 1995 \nl			
358 & \nodata & \nodata & 6.2 & 47 & 58.2 & August 5, 1995 \nl			
362 & \nodata & \nodata & 7.6 & 47 & 56.6 & August 20, 1996 \nl			
\tablebreak
381 & \nodata & \nodata & 3.1 & 47 & 52.2 & August 5, 1995 \nl			
406 & \nodata & \nodata & 3.5 & 47 & 43.5 & August 5, 1995, August 20, 1996 \nl			
577 & \nodata & \nodata & 53.8\tablenotemark{i} & 51 & 41\phantom{.}\phn & July 5, 1996 \nl
578 & \nodata & \nodata & 57.8\tablenotemark{j} & 48 & 47\phantom{.}\phn & July 5, 1996  \nl
\enddata
\tablenotetext{a}{Nagata et al.\ (1990).}
\tablenotetext{b}{Moneti, Glass, \& Moorwood (1994).}
\tablenotetext{c}{Seconds in right acsension offset from 17$^h$ 43$^m$.}
\tablenotetext{d}{Minutes and seconds of arc in declination from $-$28$^{\arcdeg}$.}
\tablenotetext{e}{The ``Pistol Star'' (Figer et al.\ 1998b). 
``Pistol Source A'' in Cotera et al.\ (1996). Object \#25 in Nagata et al.\ (1993).}
\tablenotetext{f}{Object \#26 in Nagata et al.\ (1993).}
\tablenotetext{g}{Object \#24 in Nagata et al.\ (1993).}
\tablenotetext{h}{``Pistol Source B'' in Cotera et al.\ (1996).}
\tablenotetext{i}{This star falls outside of our K$^{\prime}$-band image, so it does
not have an identification number. Coordinates are for object \#21 in Nagata et al.\ (1993). 
R.A. is in seconds of offset from 17$^h$ 42$^m$.}
\tablenotetext{j}{This star falls outside of our K$^{\prime}$-band image, so it does
not have an identification number. Coordinates are for object \#42 in Nagata et al.\ (1993). 
R.A. is in seconds of offset from 17$^h$ 42$^m$.}
\tablecomments{The Quintuplet-proper members have ``GCS'' designations in column 2. 
Coordinates are in equinox 1950.0 and are estimated from our images, 
based upon the coordinates for ``q3'' in Nagata et al.\ (1993). The ``Dates Observed'' column refers to
when spectra were obtained.}
\end{deluxetable}

\small
\clearpage
\begin{deluxetable}{rrrrrrrrrrrrrr}
\tiny
\tablecolumns{13}
\tablewidth{0pt}
\tablecaption{Photometry}
\tablehead{\colhead{ID\#} & \colhead{J}    &  \colhead{H1}   & \colhead{H2} &  \colhead{$\Delta$H} &
\colhead{K1} & \colhead{K2} & \colhead{$\Delta$K} &
\colhead{L} & 
\colhead{J$-$H} & \colhead{H1$-$K1} & \colhead{H2$-$K2} & \colhead{$\Delta$(H$-$K)} & \colhead{K$-$L} }
\startdata
76	&	16.62	&	13.50	&	13.51	&	0.01	&	11.44	&	11.45	&	0.01	&	8.80	&	3.11	&	2.06	&	2.05	&	-0.00	&	2.65	\nl
134	&	12.08	&	9.08	&	9.33	&	0.25	&	7.14	&	7.39	&	0.24	&	5.93	&	2.75	&	1.94	&	1.94	&	0.00	&	1.45	\nl
151	&	17.54	&	13.29	&	13.39	&	0.10	&	10.41	&	10.47	&	0.06	&	8.21	&	4.15	&	2.88	&	2.92	&	0.04	&	2.26	\nl
157	&	12.81	&	11.81	&	11.73	&	-0.08	&	10.38	&	10.31	&	-0.07	&	\nodata	&	1.08	&	1.43	&	1.42	&	-0.02	&	\nodata	\nl
178	&	17.17	&	13.71	&	13.63	&	-0.08	&	11.86	&	11.84	&	-0.02	&	10.49	&	3.54	&	1.85	&	1.79	&	-0.05	&	1.35	\nl
192	&	13.39	&	\nodata	&	9.68	&	\nodata	&	\nodata	&	7.74	&	\nodata	&	6.33	&	3.71	&	\nodata	&	1.94	&	\nodata	&	1.41	\nl
197	&	\nodata	&	14.23	&	\nodata	&	\nodata	&	13.66	&	\nodata	&	\nodata	&	\nodata	&	\nodata	&	0.57	&	\nodata	&	\nodata	&	\nodata	\nl
211	&	15.11	&	10.67	&	10.68	&	0.01	&	7.66	&	7.12	&	-0.54	&	3.63	&	4.43	&	3.01	&	3.55	&	0.54	&	3.50	\nl
231	&	14.00	&	9.84	&	9.62	&	-0.21	&	\nodata	&	6.27	&	\nodata	&	2.96	&	4.37	&	\nodata	&	3.35	&	\nodata	&	3.32	\nl
235	&	14.41	&	11.21	&	11.44	&	0.23	&	9.40	&	9.79	&	0.39	&	8.44	&	2.97	&	1.80	&	1.65	&	-0.16	&	1.35	\nl
240	&	14.19	&	10.86	&	11.00	&	0.14	&	9.42	&	9.10	&	-0.32	&	7.81	&	3.19	&	1.44	&	1.90	&	0.46	&	1.29	\nl
241	&	13.57	&	10.61	&	10.60	&	-0.01	&	8.66	&	8.85	&	0.19	&	6.83	&	2.97	&	1.95	&	1.75	&	-0.20	&	2.02	\nl
242	&	14.85	&	12.61	&	12.63	&	0.02	&	10.44	&	10.88	&	0.44	&	8.70	&	2.22	&	2.17	&	1.75	&	-0.42	&	2.18	\nl
243	&	15.87	&	11.12	&	11.21	&	0.08	&	\nodata	&	7.53	&	\nodata	&	4.10	&	4.66	&	\nodata	&	3.68	&	\nodata	&	3.43	\nl
250	&	15.49	&	11.69	&	11.74	&	0.05	&	9.38	&	9.24	&	-0.14	&	6.89	&	3.75	&	2.32	&	2.50	&	0.19	&	2.35	\nl
251	&	14.67	&	10.54	&	10.78	&	0.24	&	\nodata	&	7.61	&	\nodata	&	4.66	&	3.89	&	\nodata	&	3.17	&	\nodata	&	2.95	\nl
252	&	\nodata	&	13.05	&	\nodata	&	\nodata	&	9.72	&	\nodata	&	\nodata	&	\nodata	&	\nodata	&	3.33	&	\nodata	&	\nodata	&	\nodata	\nl
256	&	14.77	&	11.93	&	11.99	&	0.06	&	10.40	&	10.37	&	-0.03	&	\nodata	&	2.78	&	1.52	&	1.62	&	0.10	&	\nodata	\nl
257	&	13.41	&	\nodata	&	10.25	&	\nodata	&	\nodata	&	8.69	&	\nodata	&	7.57	&	3.16	&	\nodata	&	1.56	&	\nodata	&	1.12	\nl
258	&	\nodata	&	13.00	&	13.29	&	0.29	&	8.95	&	8.98	&	0.03	&	4.65	&	\nodata	&	4.05	&	4.31	&	0.26	&	4.33	\nl
269	&	16.81	&	12.79	&	12.79	&	-0.01	&	10.68	&	10.67	&	-0.01	&	9.57	&	4.02	&	2.11	&	2.11	&	0.01	&	1.10	\nl
270N	&	14.62	&	\nodata	&	10.75	&	\nodata	&	\nodata	&	7.81	&	\nodata	&	5.50	&	3.87	&	\nodata	&	2.94	&	\nodata	&	2.31	\nl
270S	&	\nodata	&	\nodata	&	\nodata	&	\nodata	&	\nodata	&	\nodata	&	\nodata	&	\nodata	&	\nodata	&	\nodata	&	\nodata	&	\nodata	&	\nodata	\nl
274	&	14.82	&	11.93	&	11.91	&	-0.03	&	10.41	&	10.42	&	0.01	&	9.06	&	2.92	&	1.52	&	1.49	&	-0.03	&	1.36	\nl
276	&	14.49	&	12.53	&	12.36	&	-0.17	&	10.80	&	10.82	&	0.01	&	9.97	&	2.13	&	1.73	&	1.54	&	-0.18	&	0.85	\nl
278	&	\nodata	&	10.63	&	\nodata	&	\nodata	&	9.29	&	\nodata	&	\nodata	&	\nodata	&	\nodata	&	1.34	&	\nodata	&	\nodata	&	\nodata	\nl
301	&	15.46	&	12.30	&	12.34	&	0.03	&	10.65	&	10.57	&	-08	&	9.39	&	3.13	&	1.65	&	1.76	&	0.11	&	1.18	\nl
307	&	13.92	&	10.81	&	11.03	&	0.22	&	9.38	&	9.31	&	-0.07	&	8.19	&	2.89	&	1.43	&	1.72	&	0.29	&	1.12	\nl
309	&	16.71	&	13.52	&	13.37	&	-0.15	&	11.60	&	11.44	&	-0.16	&	9.88	&	3.34	&	1.92	&	1.93	&	0.01	&	1.56	\nl
311	&	16.07	&	12.83	&	12.83	&	-0.01	&	11.12	&	11.11	&	-0.00	&	10.03	&	3.24	&	1.72	&	1.71	&	-0.00	&	1.09	\nl
320	&	15.58	&	12.44	&	12.35	&	-0.09	&	10.54	&	10.46	&	-0.08	&	8.97	&	3.23	&	1.90	&	1.89	&	-0.01	&	1.49	\nl
344	&	16.90	&	13.16	&	13.12	&	-0.04	&	11.32	&	11.23	&	-0.09	&	9.92	&	3.78	&	1.84	&	1.89	&	0.06	&	1.30	\nl
353E	&	\nodata	&	13.12	&	\nodata	&	\nodata	&	11.53	&	\nodata	&	\nodata	&	\nodata	&	\nodata	&	1.59	&	\nodata	&	\nodata	&	\nodata	\nl
358	&	15.98	&	12.27	&	12.37	&	0.10	&	10.29	&	10.26	&	-0.03	&	8.79	&	3.61	&	1.98	&	2.11	&	0.13	&	1.47	\nl
362	&	12.56	&	9.85	&	9.43	&	-0.41	&	7.99	&	7.23	&	-0.75	&	5.70	&	3.12	&	1.86	&	2.20	&	0.34	&	1.54	\nl
381	&	14.77	&	11.51	&	11.53	&	0.02	&	9.94	&	9.76	&	-0.18	&	8.48	&	3.24	&	1.56	&	1.77	&	0.20	&	1.28	\nl
406	&	15.97	&	12.87	&	12.69	&	-0.18	&	11.00	&	10.93	&	-0.07	&	9.74	&	3.28	&	1.87	&	1.76	&	-0.11	&	1.19	\nl
\enddata
\tablecomments{J, H2, K2, and L photometry are from the July 4, 1996, images. H1 and K1 photometry
are from the July, 20, 1994, images. Zero-point magnitudes were determined by observing
standard stars for the J, H1, K1, and L photometry. The zero-point magnitudes for
the H2 and K2 photometry were set by minimizing $\Delta$H and $\Delta$K.
K$^{\prime}$ photometry has been
converted to K using the relation in Wainscoat \& Cowie (1992). There has been no
conversion between nbL and L. We assess an error of 0.2 magnitudes for all photometry.}
\end{deluxetable}

\clearpage
\tiny
\clearpage
\begin{deluxetable}{rrrrl}
\tablecolumns{5}
\tablewidth{300pt}
\tablecaption{Luminosity and Ionizing Flux Estimates}
\tablehead{\colhead{ID\#} &
\colhead{Sp. Type}    &  log(L/\Lsun)\tablenotemark{a} & \colhead{log(N$_{\rm Ly,c})\tablenotemark{b}$}   & \colhead{Refs.\tablenotemark{c}}}
\startdata
76 & WC9 & 5.75& 49.3 & CS96 \nl	
134 & LBV & $\geq$6.61 & $<$41.5 & F98\nl
151 & WC8 & 6.15 & 49.7 & CS96 \nl
157 & $<$B0I & 5.67 & 48.5 & P73 \nl
%178 & nebular? & \nodata & \nodata & \nodata \nl
192 & MIa & 4.90\tablenotemark{d} & \nodata & \nodata \nl
%197 & nebular? & \nodata  & \nodata & \nodata \nl
211 & DWCL? & 4.84 & 48.7 & BSD95 \nl	
231 & DWCL? & 5.23  & 48.7 & BSD95 \nl
235 & $<$WC8 & 6.49 & 50.2 & CS96 \nl
240 & WN9/Ofpe & 6.46 & 50.1 & Naj94 \nl
241 & WN9/Ofpe & 6.66 & 50.3 & Naj94 \nl
%242 & low S/N & \nodata & 0.0 & \nodata \nl
243 & DWCL? & 4.81 & 48.7 & BSD95 \nl
250 & $<$B0I & 6.08 & 48.5 & P73 \nl
251 & DWCL? & 4.61 & \nodata & \nodata \nl
%252 & low S/N & \nodata & \nodata & \nodata \nl
256 & WN9 & 6.01 & 49.5 & CS96 \nl
257 & B0I & \nodata & 48.5 & P73 \nl
258 & DWCL? & 4.45 & 48.7 & BSD95 \nl
269 & OBI & 5.54 & 48.5 & VGS96 \nl
270N & late & \nodata & \nodata & \nodata \nl	
270S & OBI & \nodata & 48.5 & VGS96 \nl	
274 & WN9 & 6.00 & 49.5 & CS96 \nl
276 & B1I-B3I & 5.48 & 46.2 & P73 \nl
278 & OBI & \nodata & 48.5 & VGS96 \nl
301 & $<$B0I & 5.56 & 48.5 & P73 \nl
307 & B1.5Ia & 6.07 & 46.5 & P73 \nl
309 & $<$WC8 & 5.72 & 49.4 & CS96 \nl
311 & B1I-B3I & 5.36 & 46.2 & P73 \nl
320 & WN9 & 5.97 & 49.5 & CS96 \nl
344 & B1I-B3I & 5.30 & 46.2 & P73 \nl
353E & WN6 & 5.68 & 49.4 & CS96 \nl		
358 & B1I-B3I & 5.70 & 46.2 & P73 \nl
362 & OBI/LBVc & 6.44\tablenotemark{e} & 48.5 & VGS96 \nl
381 & OBI & 5.86 & 48.5 & VGS96 \nl
\tablebreak
406 & B1I-B3I & 5.42 & 46.2 & P73 \nl
\enddata
\tablenotetext{a}{The luminosities were estimated assuming d = 8,000 pc, and \AK\ = 3.28.
We assume \BCK\ of: $-$2.0 (OB supergiants), $-$3.3 (WC stars), $-$2.9 (WN and WN/Ofpe stars),  
$-$3.2\markcite{crow1} (\#353E; Crowther \& Smith 1996), and $-$1.5\markcite{f98b} (LBVs; Figer et al.\ 1998b). 
For the DWCLs, 
we integrated under the blackbody fits to the dereddened energy distributions. We 
assume \AK\ = 2.7 for the QPMs (see text).}
\tablenotetext{b}{Ionizing fluxes have been taken directly from the references, except for the
WR stars, where the following prescription has been used. N$_{\rm Ly, \, c}$ = 
q$_{0}$ L / ($\sigma$ T$^4$), where q$_{0}$ is taken from the reference, and is estimated
as 24.9 for \#353E, 23.6 for the WNL stars, 24.2 for the WC8 and WC9 stars, and 24.8 for the $<$WC8 stars; 
we assume temperatures of 55,000 K for \#353E, 30,000 for the WNL stars, 42,000 K for the WC9 star, 48,500 for
the WC8 star, and 55,000 K for the $<$WC8 stars.} 
\tablenotetext{c}{References for ionizing flux estimate: BSD95, Blum, Sellgren \& DePoy (1995);					
CS96,	Crowther \& Smith (1996);
F98, Figer et al.\ (1998b);
Har94,	Harris et al.\ (1994);			
MGM94,	Moneti, Glass \& Moorwood (1994);					
Nag90,	Nagata et al.\ (1990);					
Naj94,	Najarro et al.\ (1994);					
P73,	Panagia (1973);					
VGS96,	Vacca, Garmany \& Shull (1996).	}		
\tablenotetext{d}{We assume \BCK\ of 2.60 for an average M supergiant.}
\tablenotetext{e}{This estimate assumes the fainter photometry of July 20, 1994, and \BCK\ = $-$1.5,
typical of LBVs, and very conservative for stars hotter than 20,000 K.}
\tablecomments{Spectral types, luminosities, and ionizing fluxes for the target stars. The total
of the luminosities is log(L/\Lsun) = 7.50, or 7.36 without the two LBV stars stars (\#134 and \#362).
The total ionizing flux is N$_{Lyc}$ = 50.9 s$^{-1}$.}
\end{deluxetable}

\clearpage
\small
\begin{deluxetable}{lrrrr}
\tablecolumns{5}
\tablewidth{0pt}
\tablecaption{\AK\ Estimates from Color Excesses}
\tablehead{\colhead{} & \colhead{July 4, 1996}	&	\colhead{June 20, 1994}	&	\colhead{July 4, 1996}	&	\colhead{July 4, 1996}	\\
\colhead{}	&	\colhead{(J$-$H)}	&	
\colhead{(H$-$K)}	&	\colhead{(H$-$K)}	&	\colhead{(K$-$L\tablenotemark{a} \,)}}
\startdata
(B1I$-$B3I)$_{\rm observed}$	&	3.21	&	1.83	&	1.80	&			1.18	\nl
(B1I$-$B3I)$_{\rm intrinsic}$\tablenotemark{b}	&	$-$0.04	&	$-$0.03	&	$-$0.03	&			$-$0.07	\nl
E(B1I$-$B3I)	&	3.25	&	1.86	&	1.83	&			1.25	\nl
\AK\tablenotemark{c}	&	3.40$\pm0.60$	&	3.30$\pm0.18$	&	3.26$\pm0.34$	& 3.01$\pm0.50$	\nl
OBI$_{\rm observed}$	&	3.30	&	1.78	&	1.89	&		1.23	\nl
OBI$_{\rm intrinsic}$\tablenotemark{b}	&	$-$0.07	&	$-$0.04	&	$-$0.04	&			$-$0.07	\nl
E(OBI)	&	3.37	&	1.82	&	1.93	&			1.30	\nl
\AK\tablenotemark{c}	&	3.53$\pm1.21$	&	3.23$\pm0.38$	&	3.43$\pm1.12$	&3.13$\pm1.04$	\nl
\enddata
\tablenotetext{a}{There has been no conversion between nbL and L.}
\tablenotetext{b}{Values were taken from Koornneef 1983.}
\tablenotetext{c}{Assumes the extinction law of Rieke, Rieke, \& Paul (1989), i.e., \AJ/\AK = 2.73, 
\AH/\AK = 1.56, and \AL/\AK = 0.59.}
\tablecomments{The extinction has been inferred from measured color excesses for the early B supergiants alone 
and for the wider, and inclusive, sample of all OB supergiants. 
The B1I$-$B3I stars are: \#276, \#311, \#344, \#358, and
\#406. The ``OBI'' sample contains all of the B supergiants listed above and: \#269, \#278, \#362, and \#381.
The dates indicate when the measurements were made: note that there are two
determinations of H$-$K taken from two different nights. The unweighted 
average of all estimates is \AK\ = 3.28$\pm0.16$. The weighted average is 3.27$\pm0.06$. Note that 
the quoted statistical errors are likely less than systematic errors due to uncertainty in the
zero-points and the extinction law.}
\end{deluxetable}

\clearpage
\tiny
\begin{deluxetable}{lrrrrrrrr}
\tablecolumns{9}
\tablewidth{0pt}
\tablecaption{Properties of massive clusters}
\tablehead{
\colhead{} &
\colhead{Log(M1)} &
\colhead{Log(M2)} &
\colhead{Radius} &
\colhead{Log($\rho1$)} &
\colhead{Log($\rho2$)} &
\colhead{Age} &
\colhead{Log(L)} &
\colhead{Log(Q)} \\
\colhead{Cluster} &
\colhead{\Msun} &
\colhead{\Msun} &
\colhead{pc} &
\colhead{\Msun \, pc$^{-3}$} &
\colhead{\Msun \, pc$^{-3}$} &
\colhead{Myr} &
\colhead{\Lsun} & 
\colhead{s$^{-1}$} }
\startdata
Quintuplet& 3.0& 3.8& 1.0\phn& 2.4& 3.2& 3$-$6& 7.5 & 50.9 \nl
Arches\tablenotemark{a}& 3.7& 4.3& 0.19 & 5.2& 5.8& 1$-$2& 8.0& 51.0 \nl
Center\tablenotemark{b}& 3.0& 4.0& 0.23& 4.6& 5.6& 3$-$7& 7.3& 50.5 \nl
NGC 3603\tablenotemark{c}& 3.1& 3.7& 0.23& 4.3& 5.0& 2.5& 7.3& 51.1\nl
Trapezium\tablenotemark{d}& 1.5& \nodata& 0.05& 4.7& \nodata& 0.3 & $\approx$5\phantom{.}\phn & 48.9 \nl
R136\tablenotemark{e}& 3.4& 4.5& 1.6\phn& 2.2& 3.3& $<$1$-$2& $>$7.6& 51.9\nl
small globular (M5)\tablenotemark{f}& \nodata& 4.8& 4.0\phn& \nodata& 2.3& \nodata& \nodata& \nodata \nl
typical globular(M13)\tablenotemark{f}& \nodata& 5.5& 3.9\phn& \nodata& 3.1& \nodata& \nodata& \nodata \nl
large globular (M22)\tablenotemark{f}& \nodata& 6.8& 3.2\phn& \nodata& 4.7& \nodata& \nodata& \nodata \nl
NGC 1705-1\tablenotemark{g}& \nodata & 4.9 & 0.9\phn& \nodata & 4.4& 10$-$20& \nodata& \nodata \nl
NGC 1569-A\tablenotemark{g}& \nodata& 5.5& 1.9\phn& \nodata& 4.0& 10$-$20& \nodata& \nodata \nl
%NGC 4214-1\tablenotemark{l}& \nodata& \nodata& $\leq$ 2.5& \nodata& \nodata& 4$-$5& \nodata& \nodata \nl
\enddata
\tablenotetext{a}{Serabyn, Shupe, \& Figer (1998).}
\tablenotetext{b}{Krabbe et al.\ (1995). The mass, ``M2'' has been estimated by assuming that a total 10$^{3.5}$
stars have been formed. The age spans a range covering an initial starburst, followed by
an exponential decay in the star formation rate.}
\tablenotetext{c}{Drissen et al.\ (1995). ``M1'' was estimated by assuming that the 37 stars in Table 3 of
Moffat, Drissen, \& Shara (1994) have 15 $<$ M$_{\rm initial}$ $<$ 120 \Msun. The ``size'' is the average
projected radius for the stars in Table 3. The ``age'' is estimated by assuming that there are true
helium-burning Wolf-Rayet stars in the cluster. See Eisenhauer et al.\ (1998) for an alternate interpretation.
The luminosity has been estimated by assuming that the
stars in Table 3 have an average luminosity of 10$^{5.7}$ \Lsun.}
\tablenotetext{d}{McCaughrean \& Stauffer (1994). The ``radius'' comes from the region which was
probed with deep infrared imaging. ``$\rho$2'' is the stellar number density, i.e., it represents
the mass density if the average mass per star is 1 \Msun. This number
has been corrected for projection effects, i.e., volume number density $\approx$ 2 $\times$ surface number density.
``Q'' is from Kennicutt 1984.}
\tablenotetext{e}{Massey \& Hunter (1998). ``M1'' is a count of the total mass of the 29 stars in Table 3
of Massey \& Hunter, using the Chlewbowski \& Garmany (1991) calibration 
(56 \Msun\ $<$ M$_{\rm initial}$ $<$ 136 \Msun). 
The age corresponds to the young population of high mass stars. ``Q'' is from Walborn 1991.}
\tablenotetext{f}{Allen (1973).}
\tablenotetext{g}{Ho \& Filippenko (1996). The masses, ``M2,'' are inferred by measuring
velocity dispersions and invoking the virial theorem.}
\tablecomments{
``M1'' is the total cluster mass in observed stars. ``M2'' is the total cluster mass in all
stars extrapolated down to a lower-mass cutoff of 1 \Msun, assuming a Salpeter IMF slope and an
upper mass cutoff of 120 \Msun (unless otherwise noted); note that the total cluster mass would be 2.5 times greater if
the lower mass cutoff is 0.1 \Msun.
``Radius'' gives the average projected separation from the centroid position. 
``$\rho1$'' is M1 divided by the volume. ``$\rho2$'' is M2 divided by the volume. In either case, 
this is probably closer to the central density than
the average density because the mass is for the whole cluster while the radius is the
average projected radius. ``Age'' is the assumed age for the cluster. ``Luminosity'' gives
the total measured luminosity for observed stars. ``Q'' is the estimated Lyman continuum
flux emitted by the cluster.}
\end{deluxetable}

\newpage

\figurenum{1}
\figcaption[quinjhk.ps]
{JHK$^{\prime}$ color composite of the Quintuplet Cluster, covering 3$\arcmin$ $\times$ 3$\arcmin$.}

\figurenum{2a}
\figcaption[quinj.ps]
{J-band image of Quintuplet Cluster, covering 3$\arcmin$ $\times$ 3$\arcmin$. }

\figurenum{2b}
\figcaption[quinh.ps]
{H-band image of Quintuplet Cluster, covering 3$\arcmin$ $\times$ 3$\arcmin$.}

\figurenum{2c}
\figcaption[quink.ps]
{K$^{\prime}$-band image of Quintuplet Cluster, covering 3$\arcmin$ $\times$ 3$\arcmin$.}

\figurenum{2d}
\figcaption[quinl.ps]
{nbL-band image of Quintuplet Cluster, covering 3$\arcmin$ $\times$ 3$\arcmin$. 
The diagonal lines are artifacts from bad pixels
in the array.}

\figurenum{3a}
\figcaption[z435k.ps]
{Alternate stretch of Figure 2c at an expanded scale with respect to Figures 1-2.}

\figurenum{3b}
\figcaption[quinstar.ps]
{Schematic diagram of the stars covered in the slit-scan. Individual stars are labeled
with number according to Table 1. The slit-scan did not include stars \#353e and \#362.}

\figurenum{4}
\figcaption[qmap.ps]
{Identified spectral types in the Quintuplet Cluster. The slit-scan area is 
shown as a rectangular box. See text for a description of how the stellar
types were determined.}

\figurenum{5}
\figcaption[wrall.ps]
{K-band spectra of WR stars in the Galaxy obtained with the same instrument and setup as was
used to obtain the target spectra in this paper (R $\approx$ 525). The identification numbers are taken from
van der Hucht et al.\ (1981). WC stars are shown in the left panels, and WN stars are shown
in the right panels. Likely emission line identifications are indicated by tick
marks.}

\figurenum{6a}
\figcaption[quinwn9.ps]
{K-band spectra of WN9 stars in the Quintuplet and WR108 (WN9). Features near 2.05 \micron, 2.315 \micron,
and 2.370 \micron, are suspect due to difficulties in correcting for atmospheric absorption; tick marks
indicate that features at these wavelengths are uncertain. All subsequent
spectra have been dereddened according to the text.}

\figurenum{6b}
\figcaption[quinwn6.ps]
{K-band spectra of \#353e (WN6) in the Quintuplet and WR115 (WN6). See caption for Figure 6a for more.}

\figurenum{6c}
\figcaption[quinwc9.ps]
{K-band spectra of \#76 (WC9) in the Quintuplet and WR121 (WC9).}

\figurenum{6d}
\figcaption[quinwc8.ps]
{K-band spectra of WC8 and ``<WC8'' stars in the Quintuplet and WR135 (WC8). We have used measurements of
the excess emission at 3.09 \micron \ion{He}{2} lines to classify the
``$<$WC8'' stars. See caption for Figure 6a for more.}

\figurenum{7}
\figcaption[quinob.ps]
{K-band spectra of OB supergiants in the Quintuplet. 
In cases where two spectra are shown, the
upper spectrum was taken on the earlier date. A curve
for a hot blackbody has been overplotted for comparison.
The ``$<$B0I'' stars have nearly
featureless spectra which can only be matched to spectra of stars earlier than B0I. 
The ``OBI'' stars have an ambiguous classification, showing a hint of absorption in 
the 2.112 \micron\ \ion{He}{1} line and the 2.166 \micron\ \ion{H}{1} line with a
small amount of excess emission in the 2.058 \micron\ \ion{He}{1} line in some cases.
The dereddened spectra are steeper than the blackbody curve in some cases, i.e., \#270S. Evidently, \AK\ = 3.2 is
too high for this star. Note the lack of CO absorption features, i.e., at 2.294 \micron,
in the spectrum for \#362 (see text). See caption for Figure 6a for more.}

\figurenum{8}
\figcaption[quinb.ps]
{K-band spectra of early B supergiants in the Quintuplet. In cases where two spectra are shown, the
upper spectrum was taken on the earlier date. 
A curve
for a hot blackbody has been overplotted for comparison.
Features at  
2.312 \micron\ and 2.370 \micron\ are due to imperfect atmospheric correction.  
The spectra for \#311 and \#344 have
low signal-to-noise. The August 20, 1996, spectrum of \#311 reveals a late-type companion
as seen in the CO absorption features longward of 2.29 \micron. See caption for Figure 6a for more.}

\figurenum{9a}
\figcaption[qpm1.ps]
{Dereddened spectral energy distribution of \#243 (\#1 in MGM94). We assume \AK\ = 2.7
because this value gives the best fit between the observations and the blackbodies.
A best-fit blackbody curve has been overplotted here and in Figures 9b-e.}

\figurenum{9b}
\figcaption[qpm2.ps]
{Dereddened spectral energy distribution of \#231 (\#2 in MGM94). See Figure 9a for more.}

\figurenum{9c}
\figcaption[qpm3.ps]
{Dereddened spectral energy distribution of \#211 (\#3 in MGM94). See Figure 9a for more.}

\figurenum{9d}
\figcaption[qpm4.ps]
{Dereddened spectral energy distribution of \#251 (\#4 in MGM94). See Figure 9a for more.}

\figurenum{9e}
\figcaption[qpm9.ps]
{Dereddened spectral energy distribution of \#258 (\#9 in MGM94). See Figure 9a for more.}

\figurenum{10}
\figcaption[quinnear.ps]
{K-band spectra of two stars near the Quintuplet. The spectra 
are well-fit by cool blackbody curves. See caption for Figure 6a for more.}

\figurenum{11}
\figcaption[isq40m2c]
{Model isochrones from the Geneva models for Z = 0.04 with ``2$\times$'' mass-loss. Isochrones
begin at 1 Myr and are spaced by 1 Myr. Data for early B supergiants 
in the Quintuplet Cluster are overplotted. The error in temperature covers the range for
B1I$-$B3I stars. The error in luminosity is dominated by errors in determining the extinction.
Assuming that the upper data points are influenced by confusion or binarity, we find a
cluster age $\approx$ 4 Myr.
}

\newpage

\begin{figure}
%this is a color plot, so it will have to be added later
\plotone{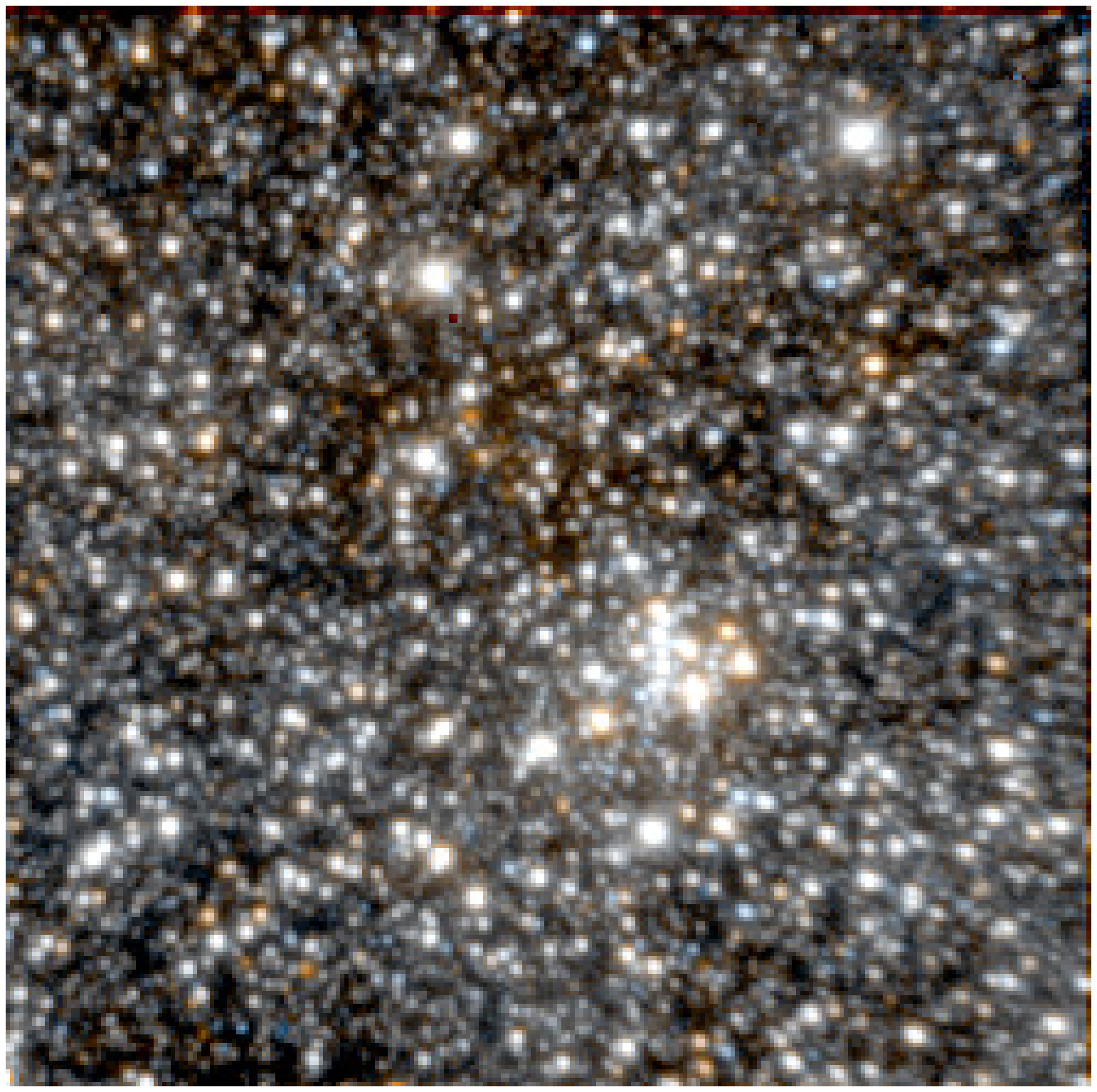}
\end{figure}

\begin{figure}
\plotone{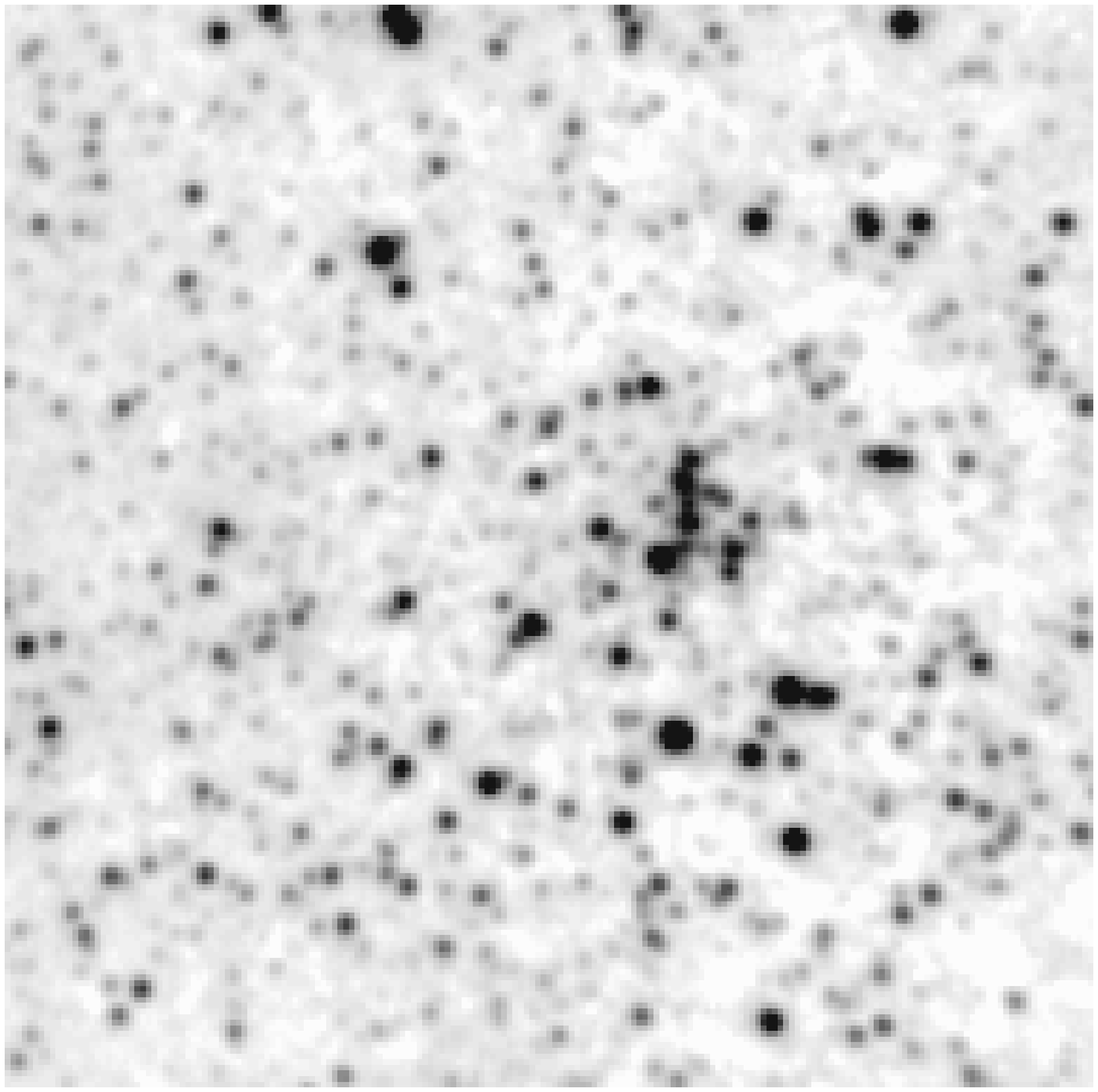}
\hspace*{4.5in} 
\vskip .2in
Figure 2a
\end{figure}

\begin{figure}
\plotone{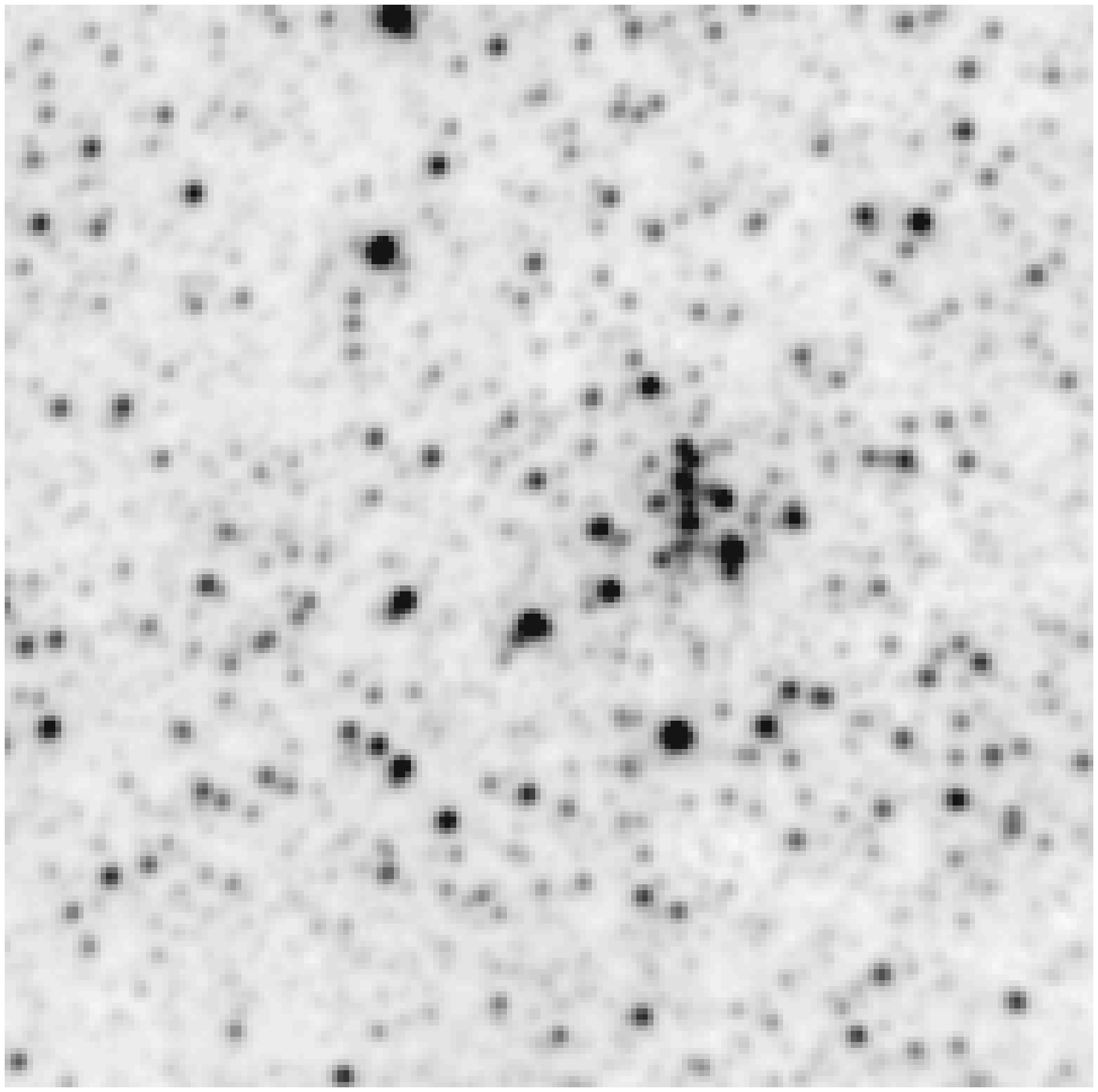}
\hspace*{4.5in} 
\vskip .2in
Figure 2b
\end{figure}

\begin{figure}
\plotone{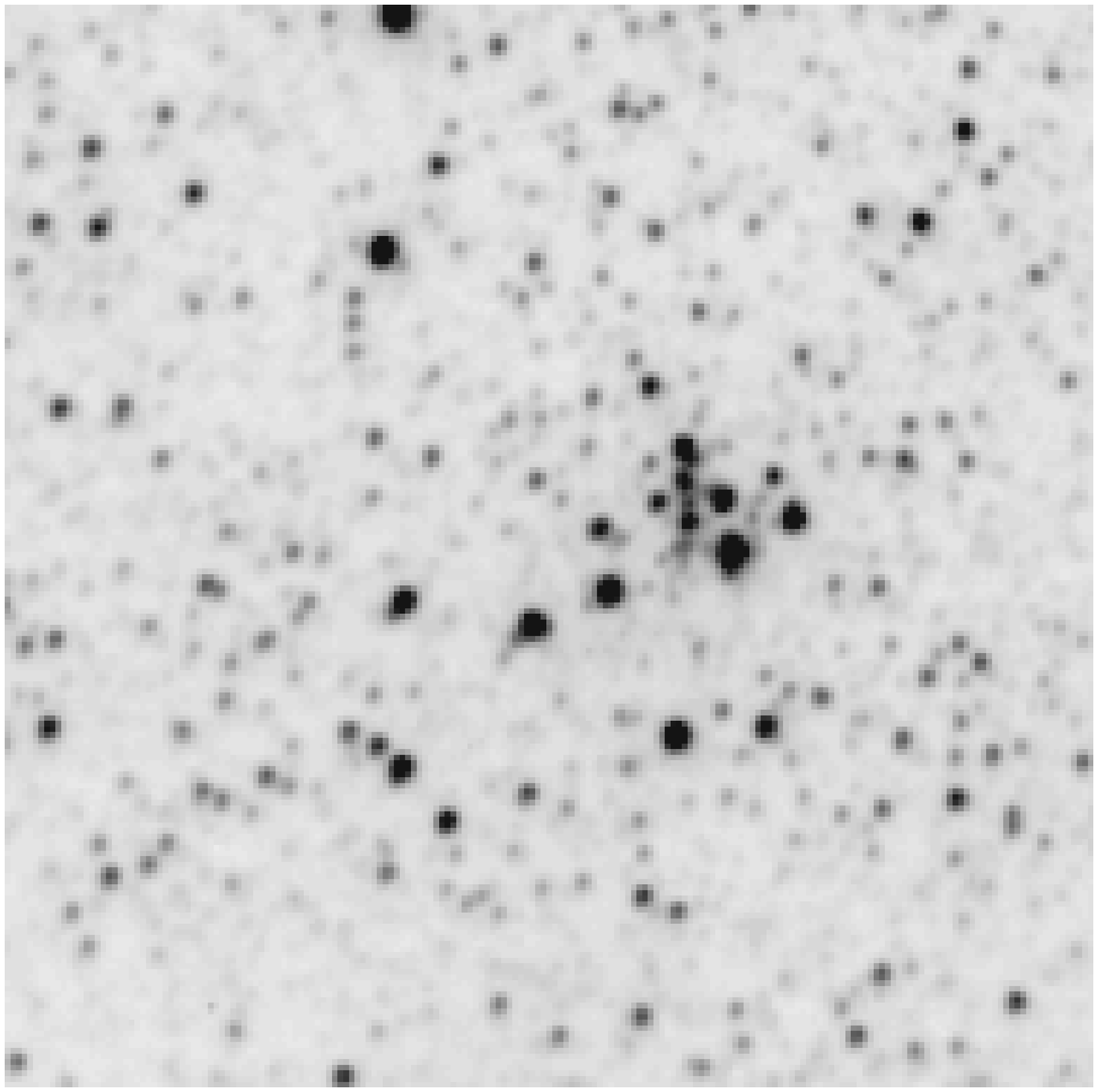}
\hspace*{4.5in} 
\vskip .2in
Figure 2c
\end{figure}

\begin{figure}
\plotone{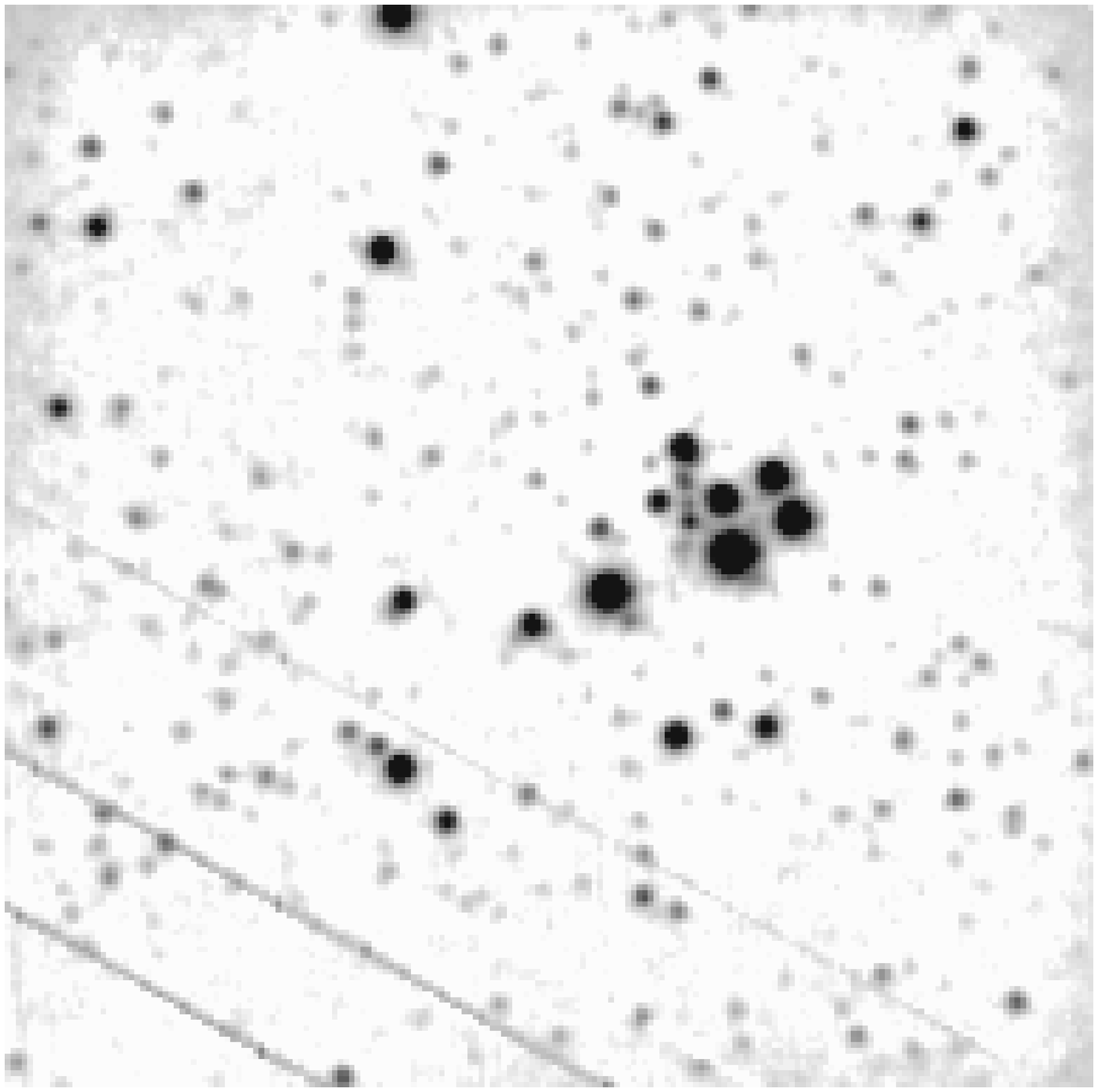}
\hspace*{4.5in} 
\vskip .2in
Figure 2d
\end{figure}

\begin{figure}
\plotone{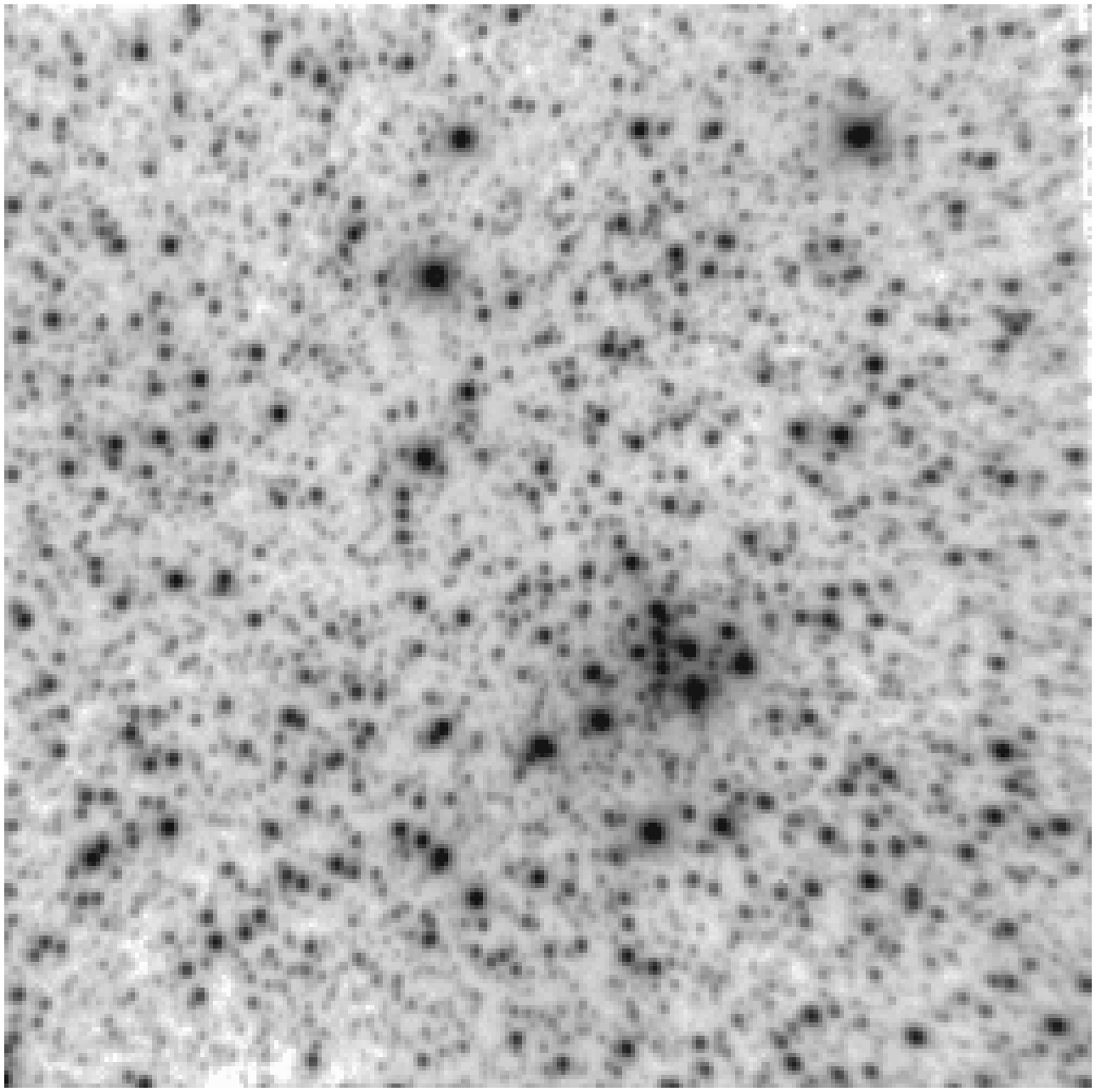}
\hspace*{4.5in} 
\vskip .2in
Figure 3a
\end{figure}

\begin{figure}
\plotone{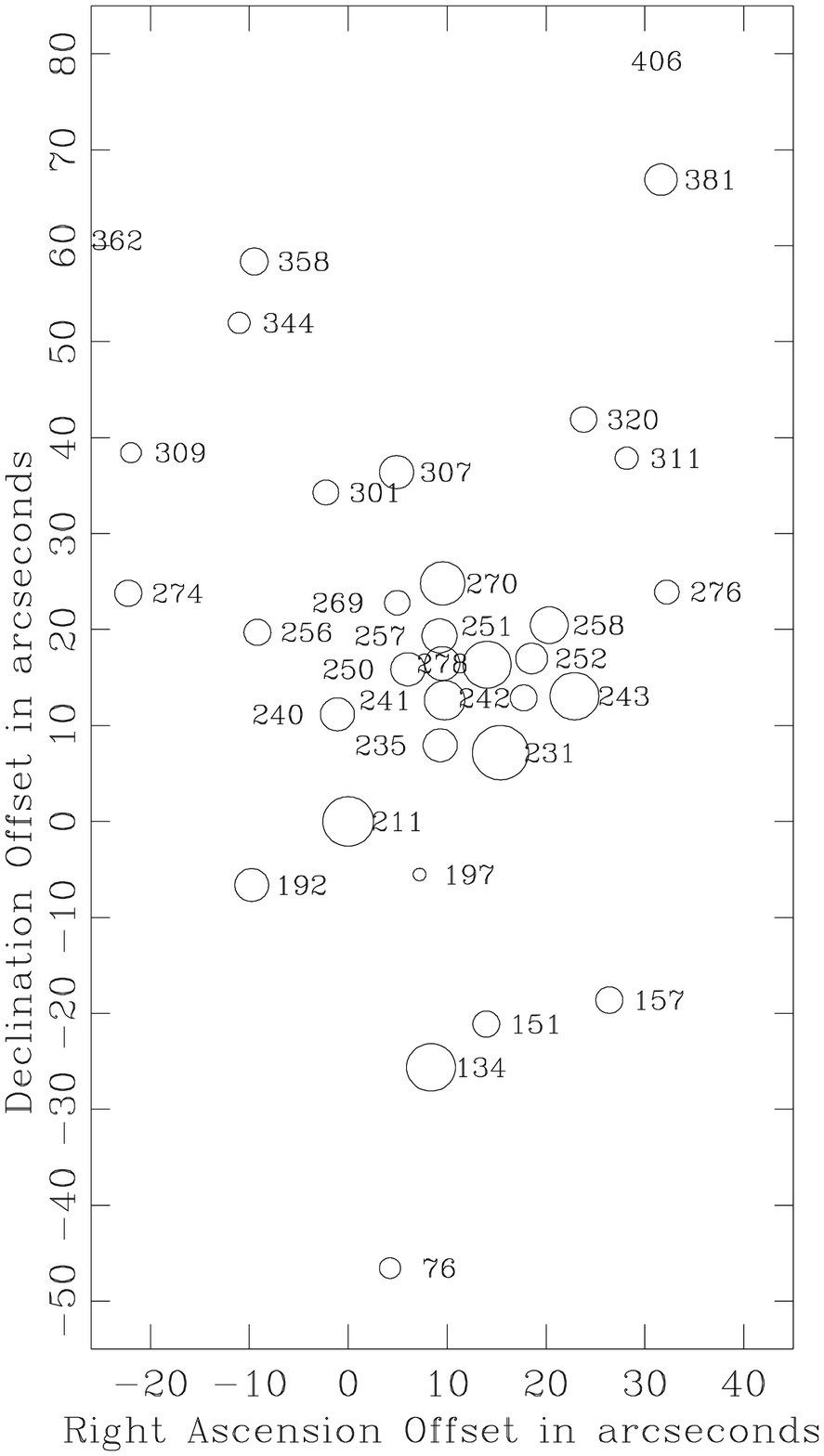}
\hspace*{4.5in} 
\vskip .2in
Figure 3b
\end{figure}

\begin{figure}
\plotone{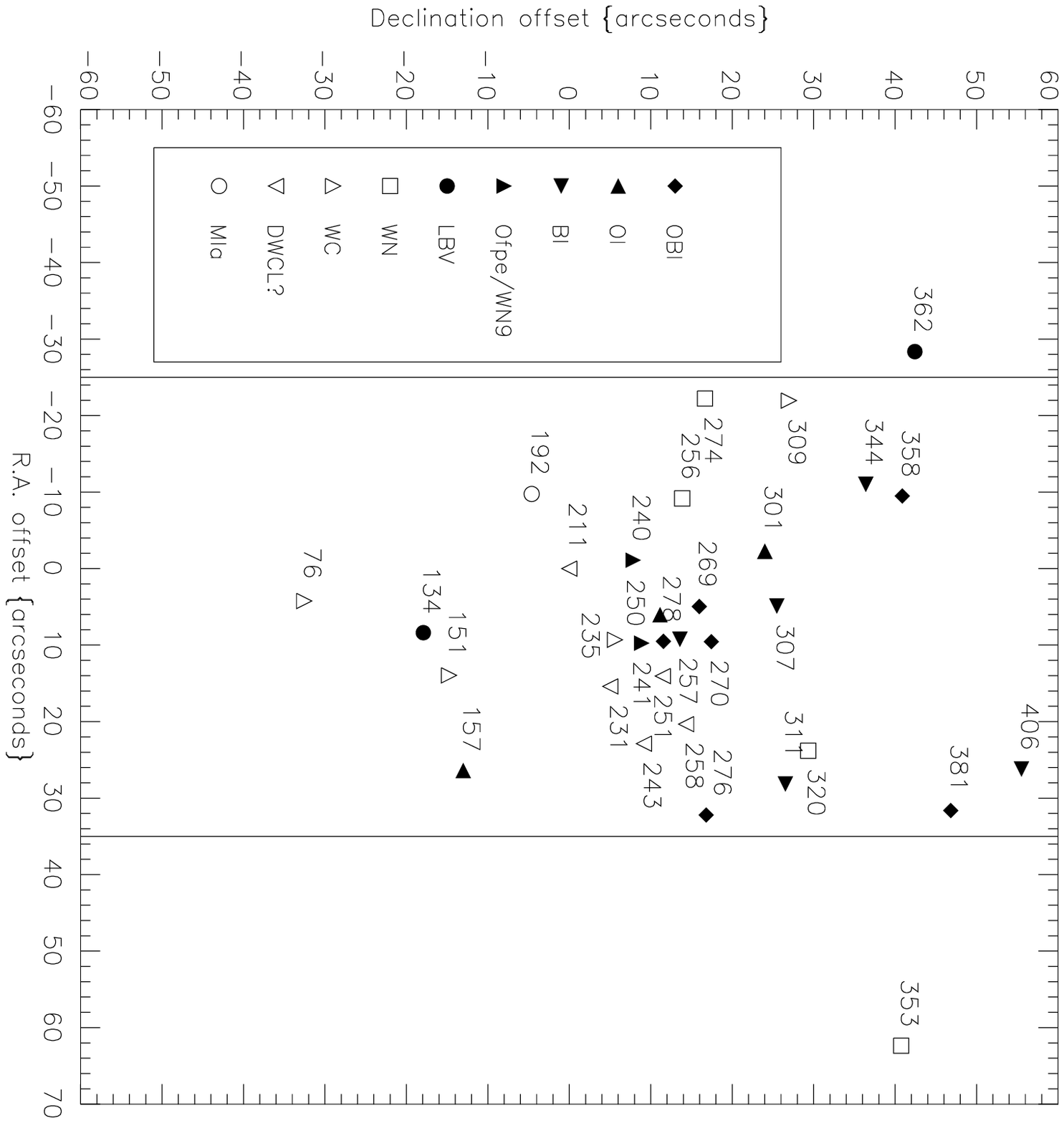}
\hspace*{4.5in} 
\vskip .2in
Figure 4
\end{figure}

\clearpage

\begin{figure}
\plotone{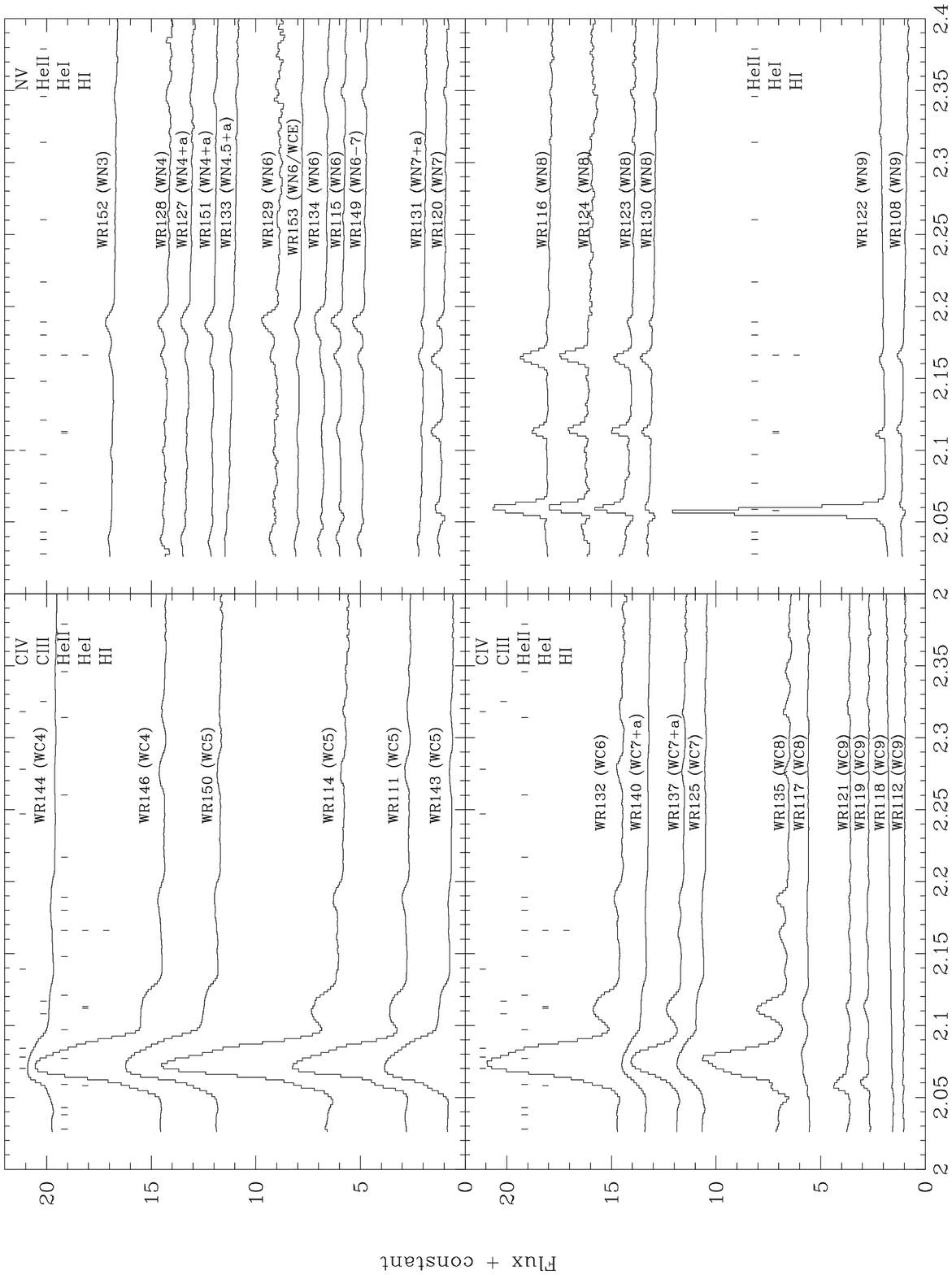}
\hspace*{4.5in} 
\vskip .2in
Figure 5
\end{figure}

\begin{figure}
\plotone{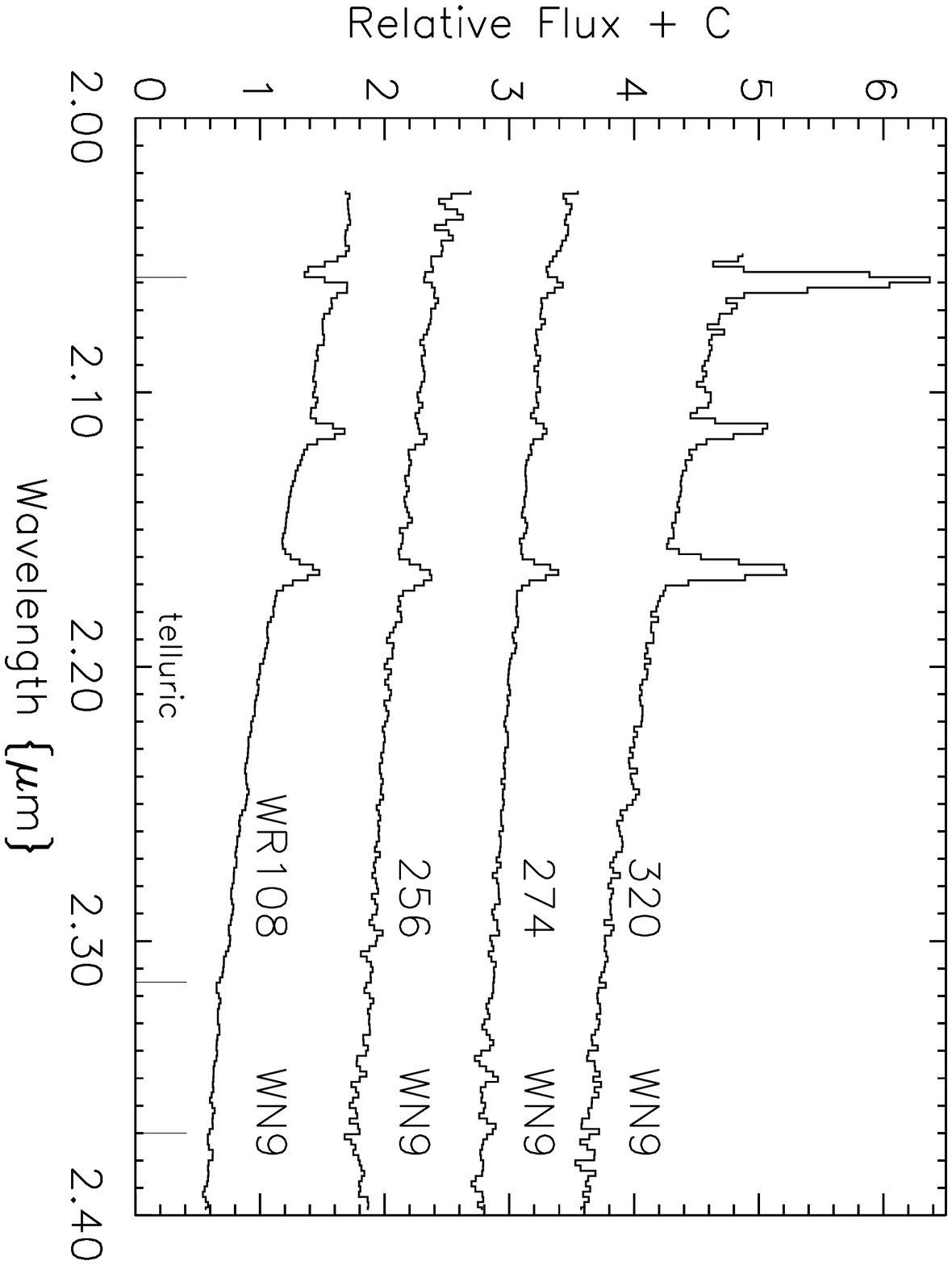}
\hspace*{4.5in} 
\vskip .2in
Figure 6a
\end{figure}

\begin{figure}
\plotone{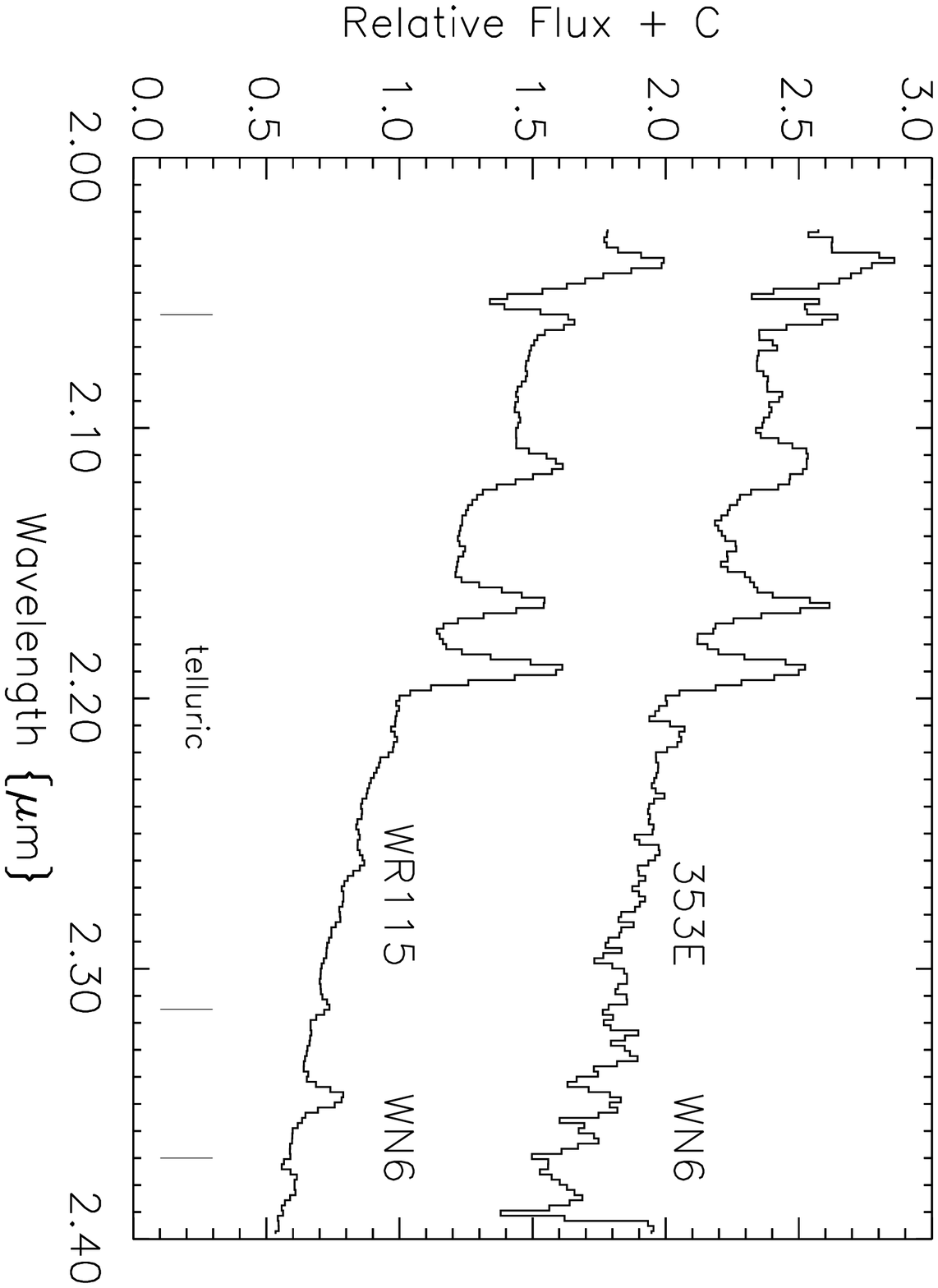}
\hspace*{4.5in} 
\vskip .2in
Figure 6b
\end{figure}

\begin{figure}
\plotone{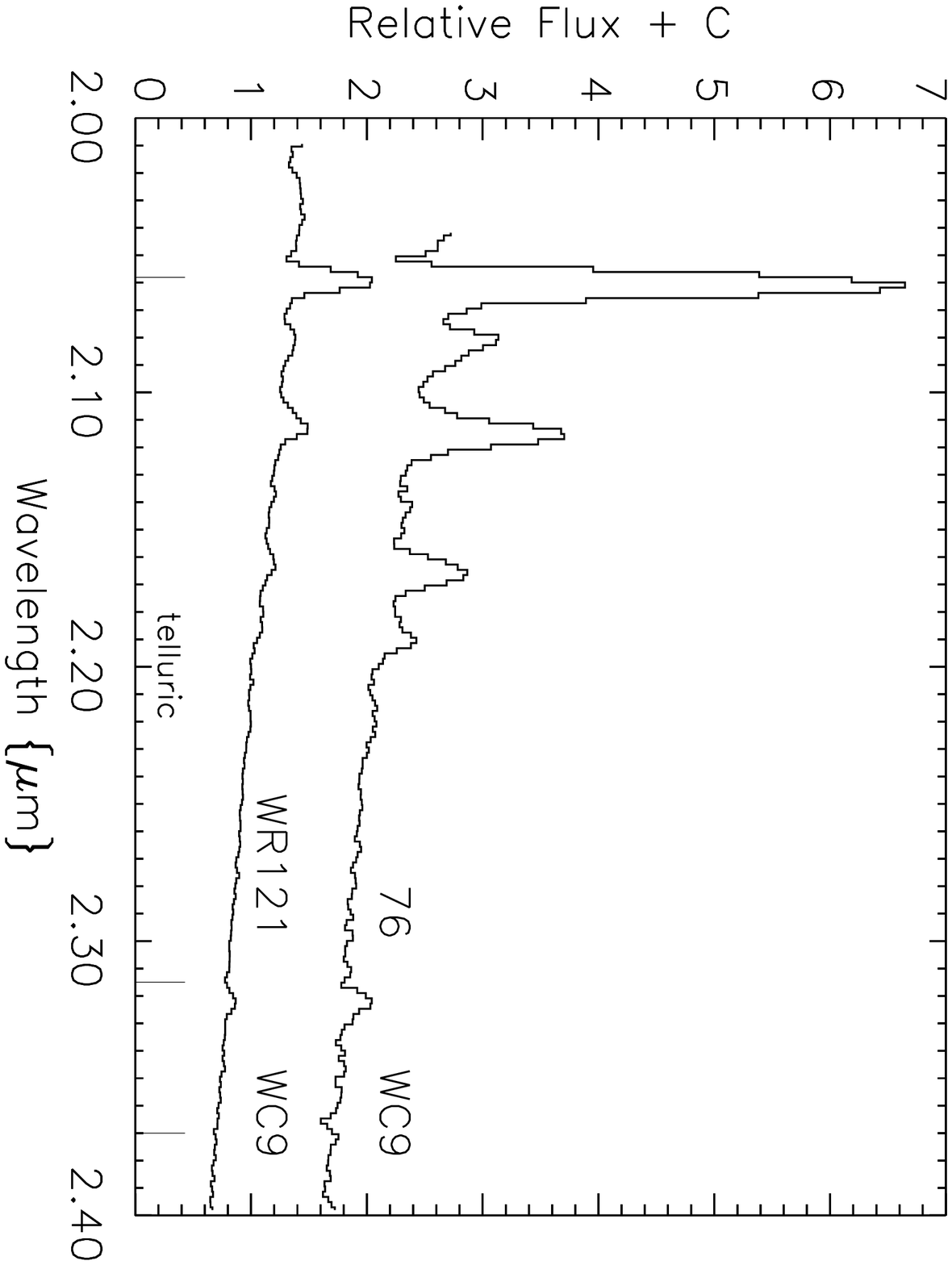}
\hspace*{4.5in} 
\vskip .2in
Figure 6c
\end{figure}

\begin{figure}
\plotone{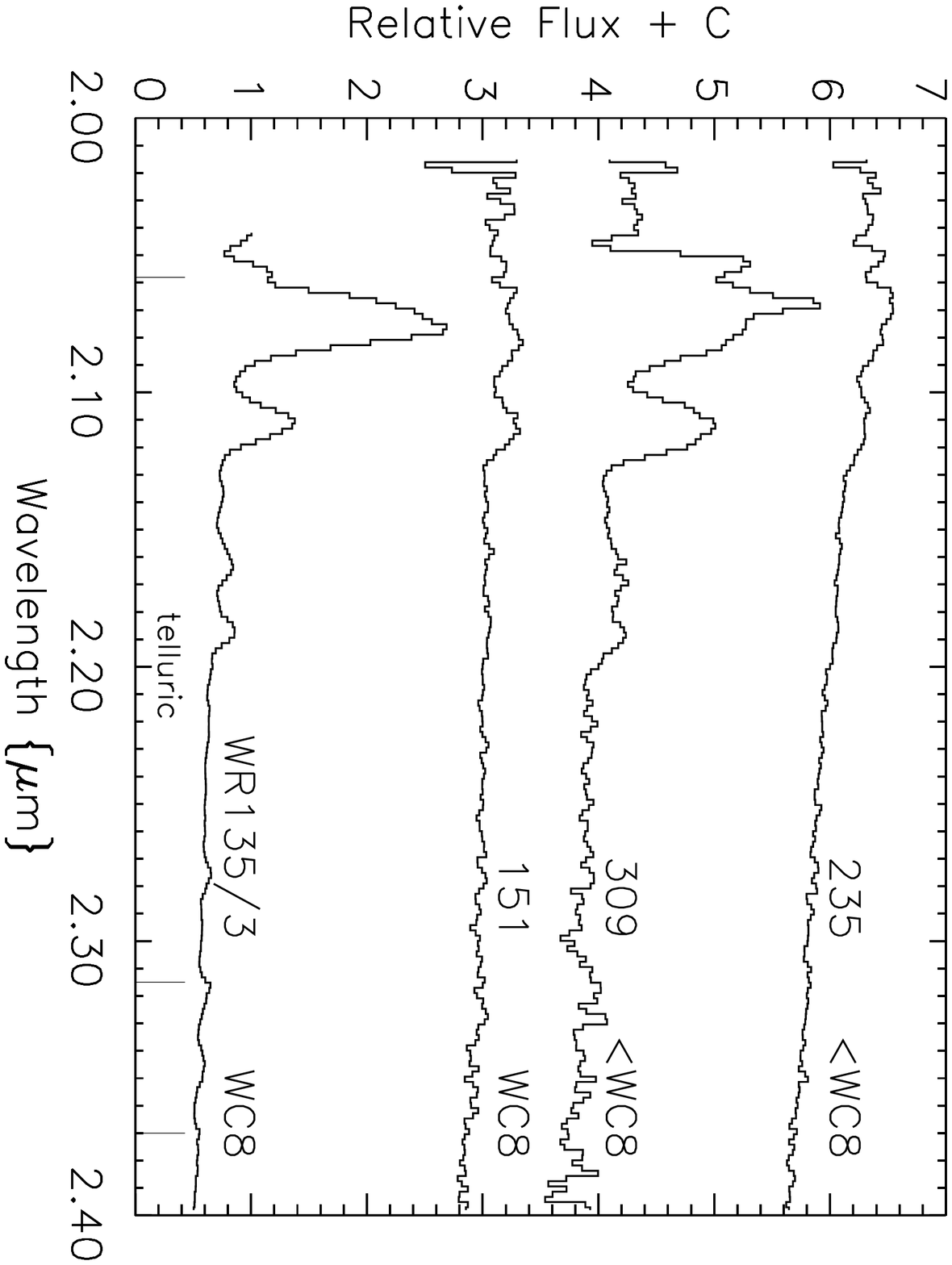}
\hspace*{4.5in} 
\vskip .2in
Figure 6d
\end{figure}

\begin{figure}
\plotone{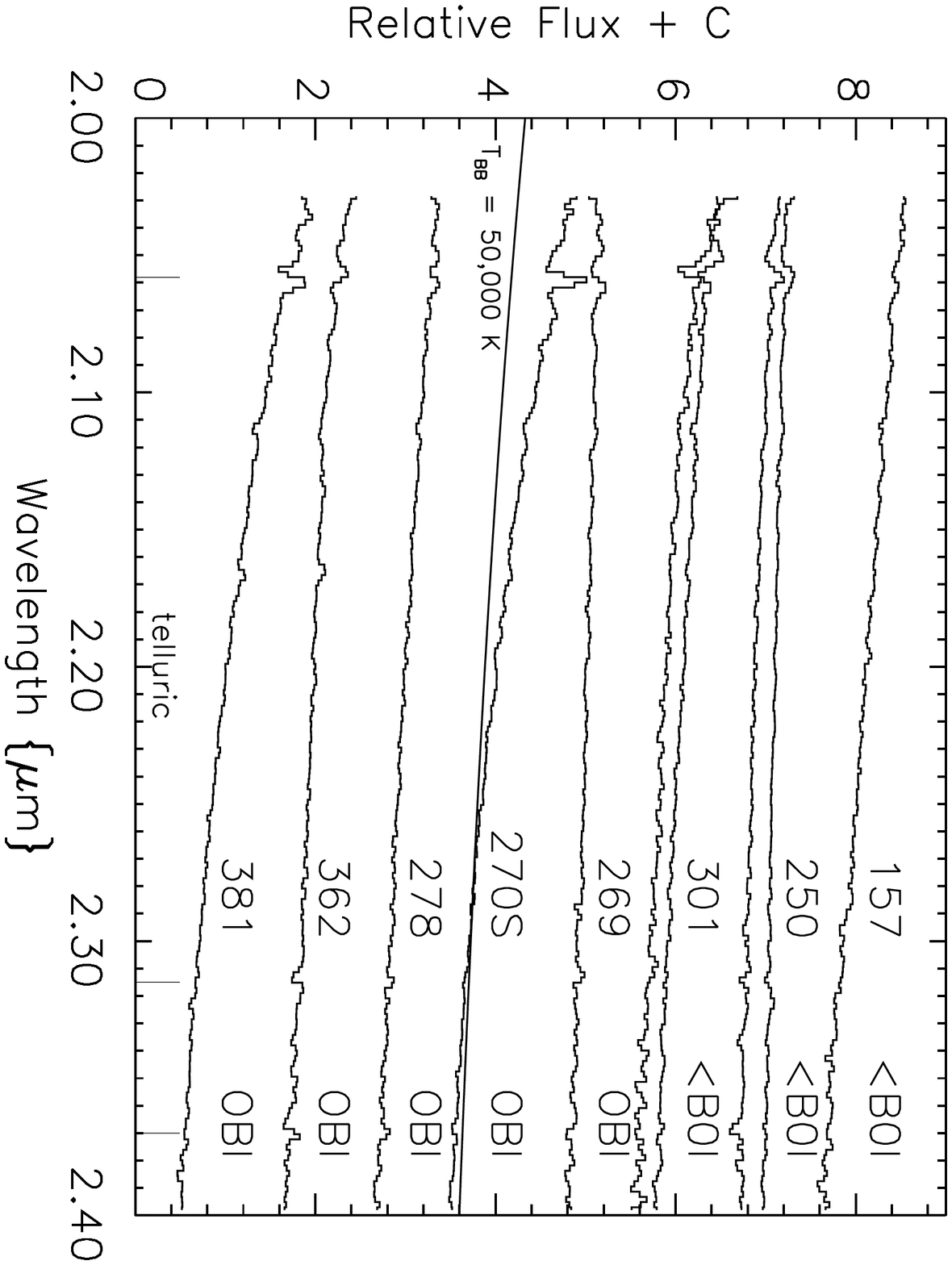}
\hspace*{4.5in} 
\vskip .2in
Figure 7
\end{figure}

\begin{figure}
\plotone{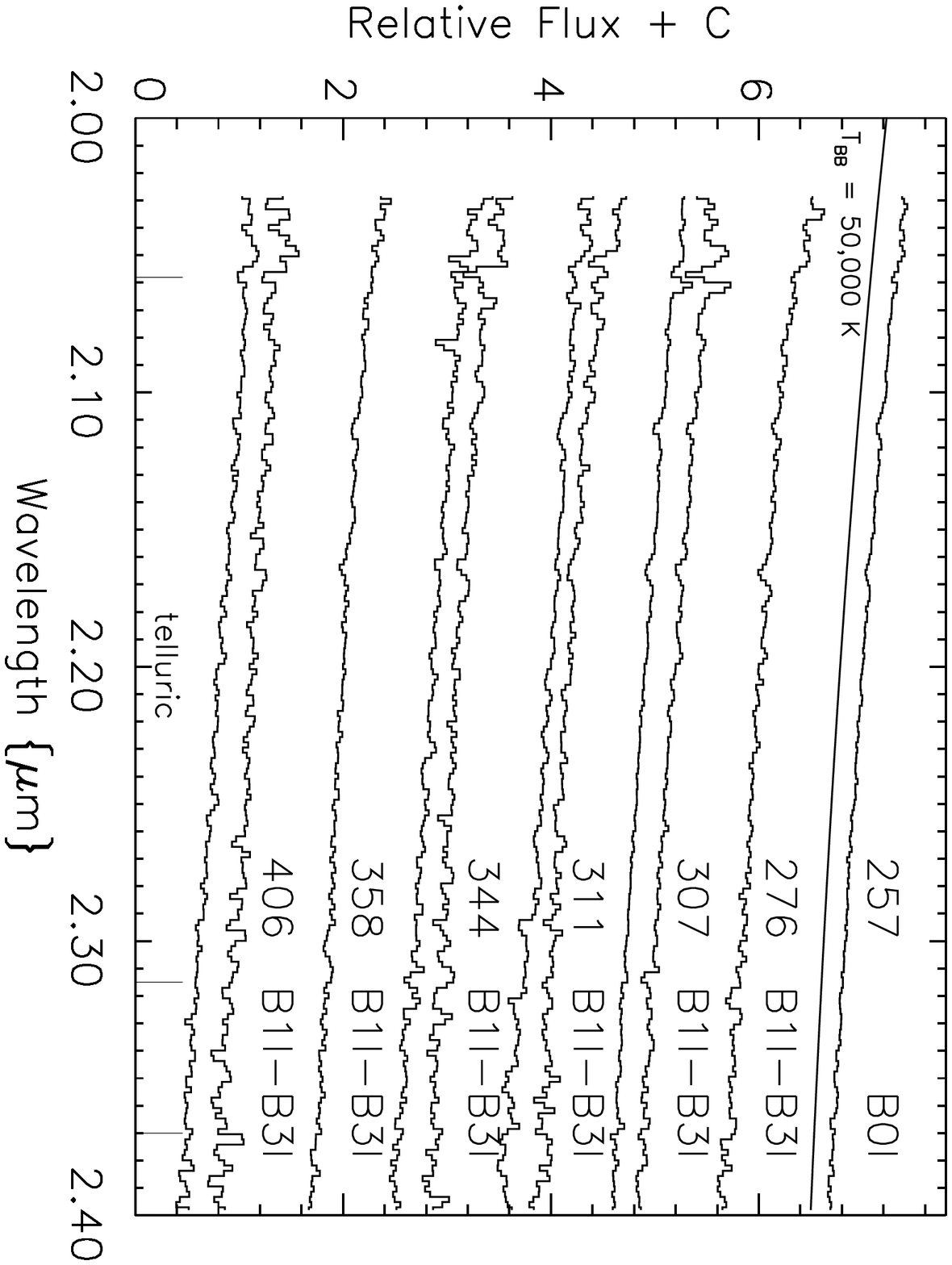}
\hspace*{4.5in} 
\vskip .2in
Figure 8
\end{figure}

\clearpage

\begin{figure}
\plotone{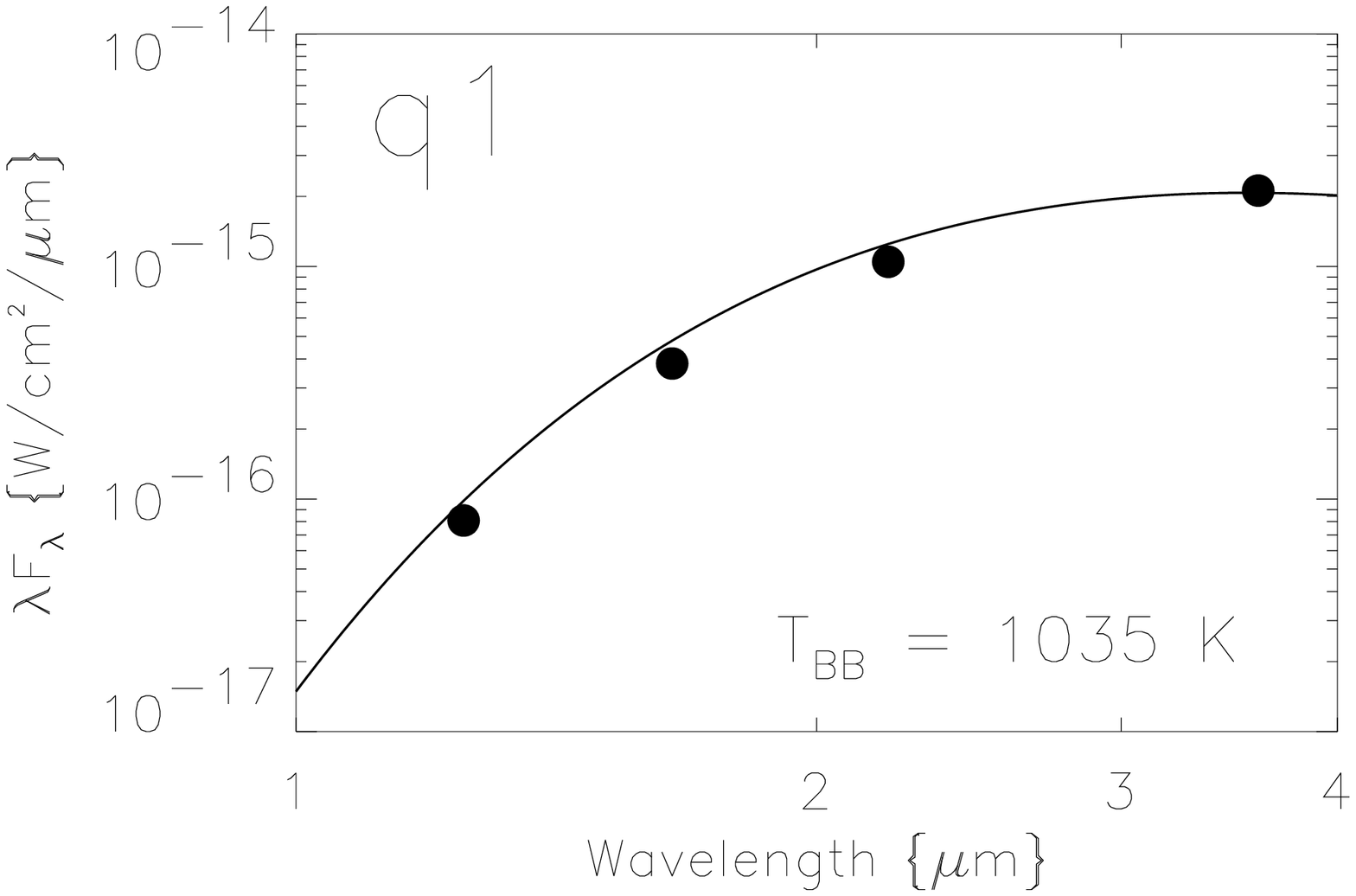}
\hspace*{4.5in} 
\vskip .2in
Figure 9a
\end{figure}

\begin{figure}
\plotone{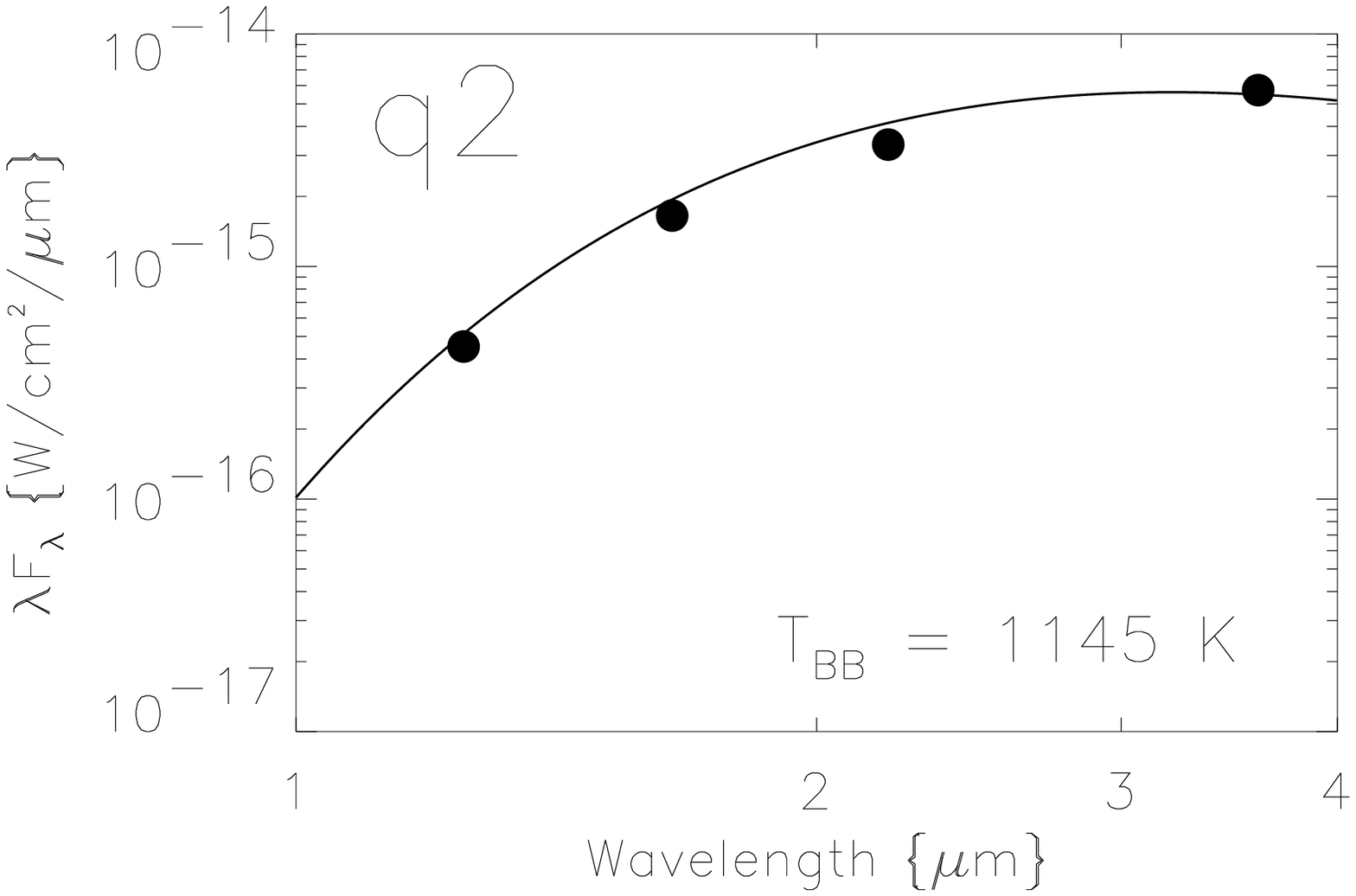}
\hspace*{4.5in} 
\vskip .2in
Figure 9b
\end{figure}

\begin{figure}
\plotone{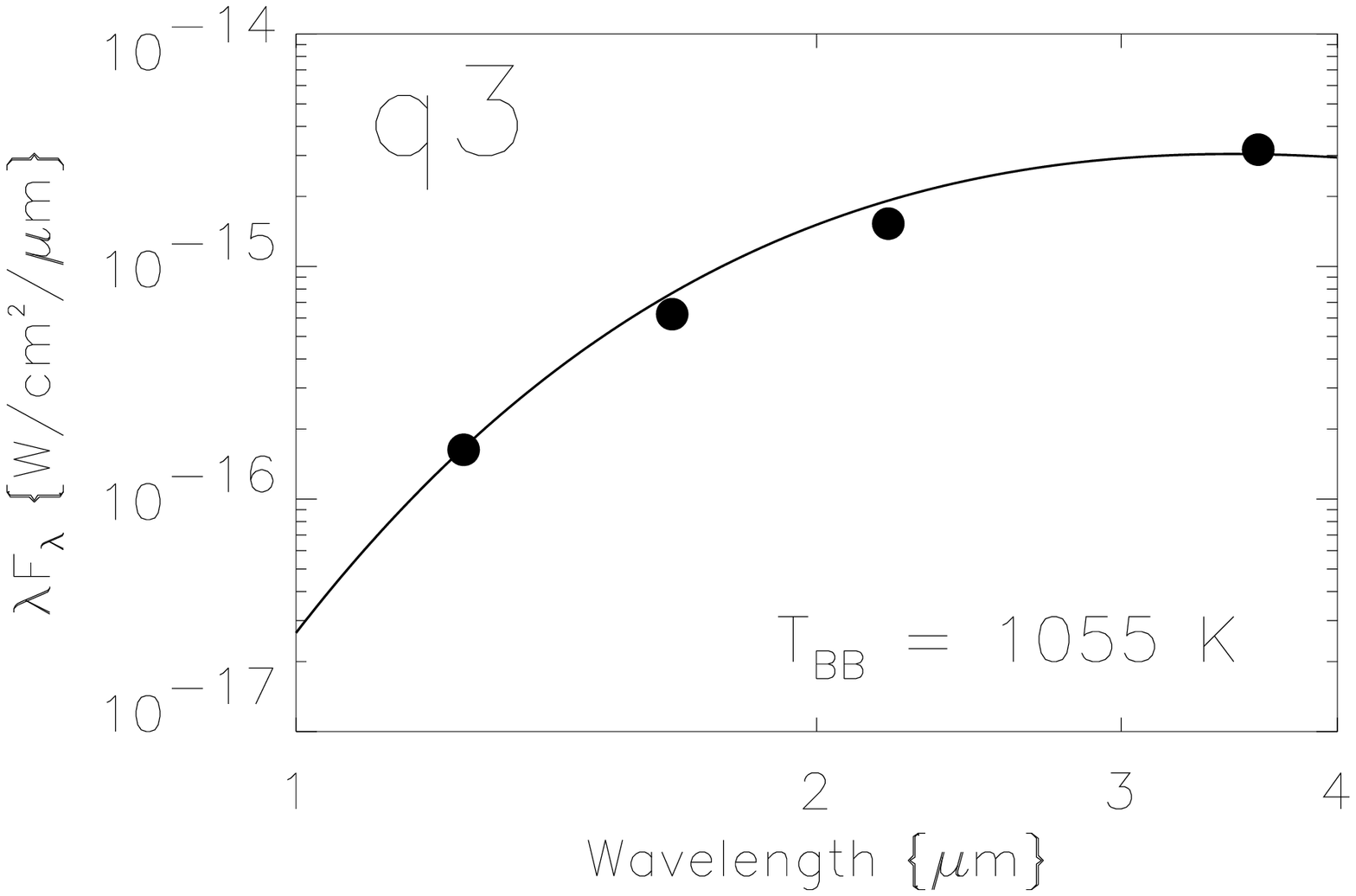}
\hspace*{4.5in} 
\vskip .2in
Figure 9c
\end{figure}

\begin{figure}
\plotone{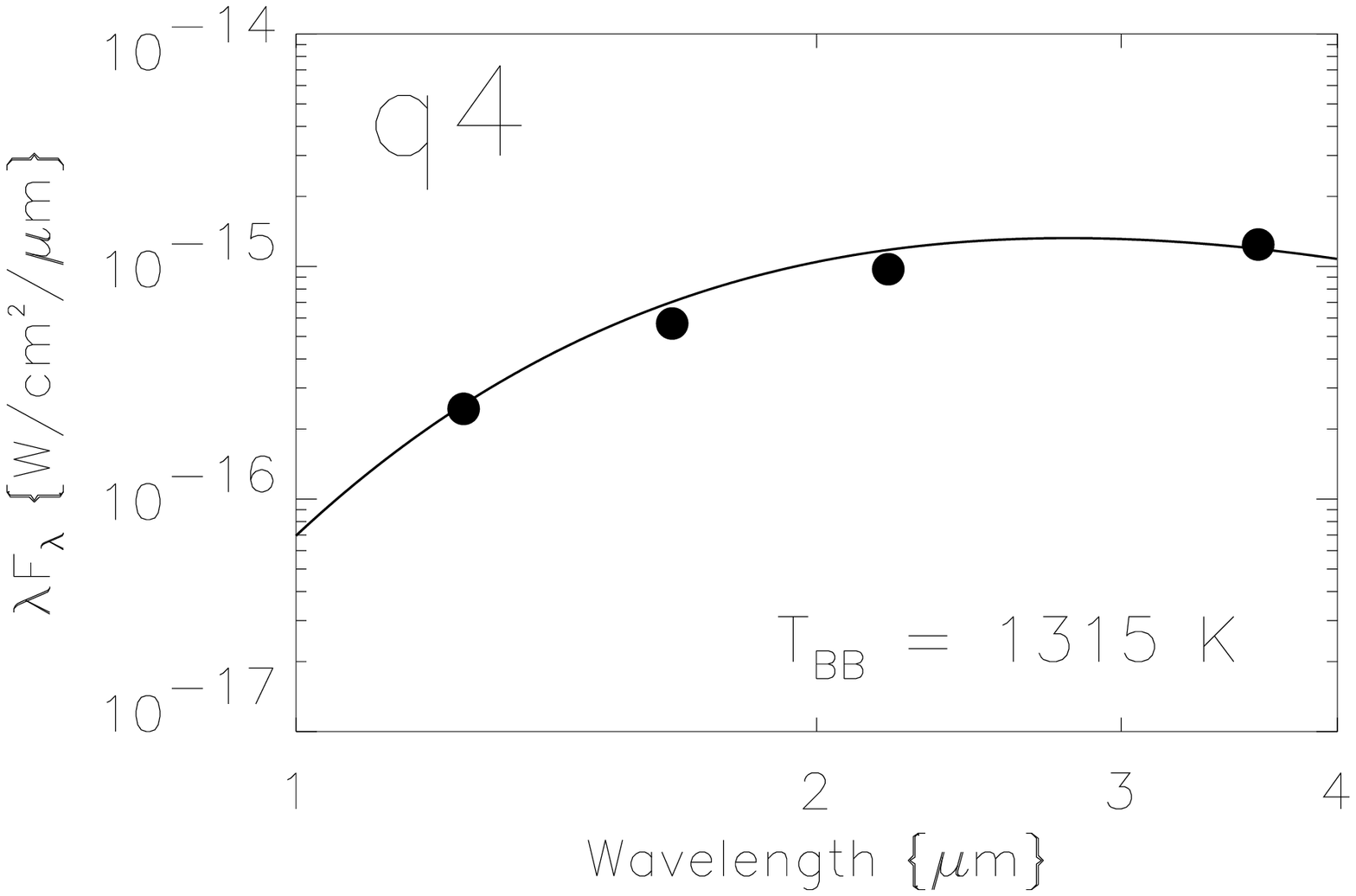}
\hspace*{4.5in} 
\vskip .2in
Figure 9d
\end{figure}

\begin{figure}
\plotone{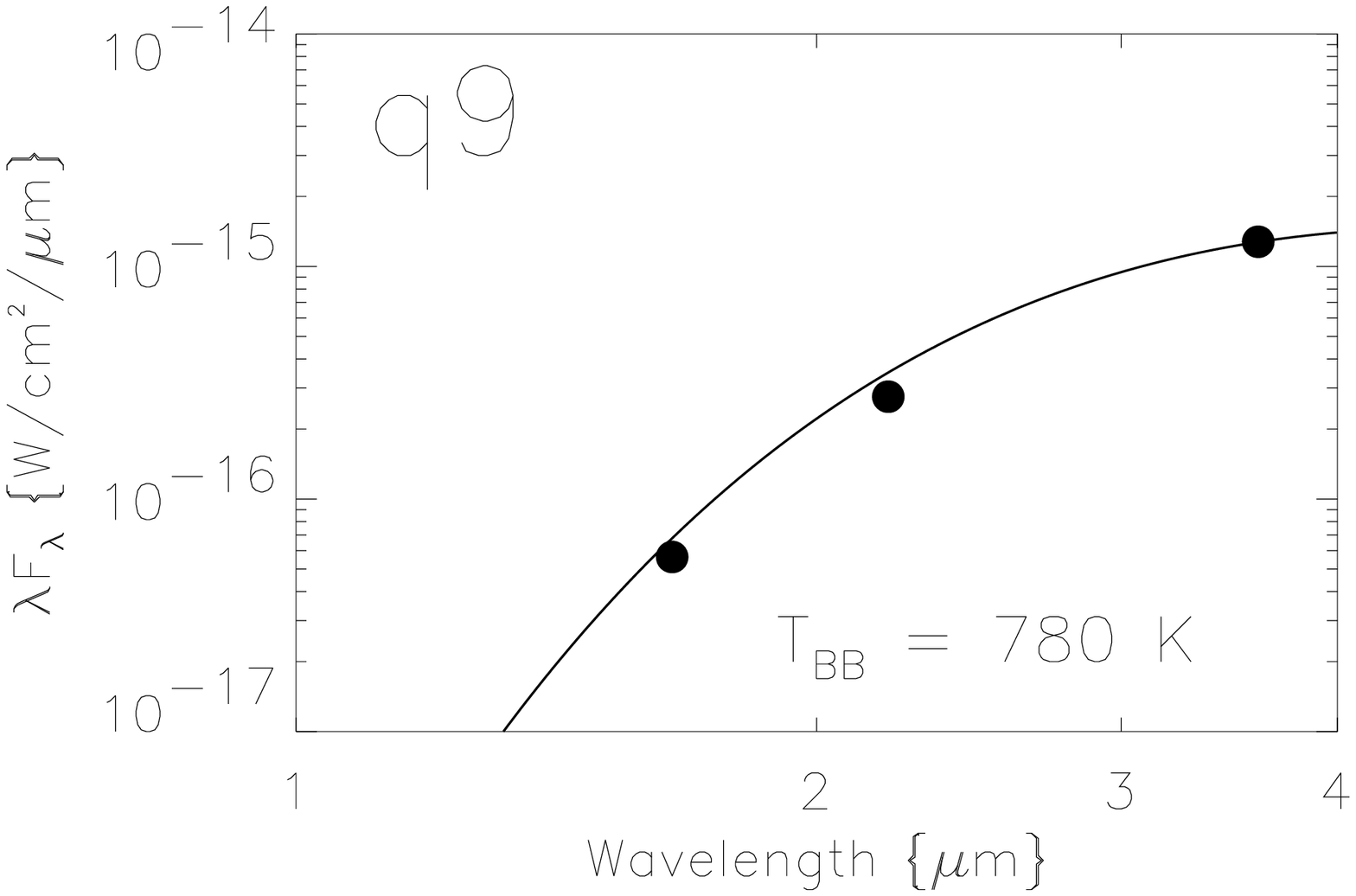}
\hspace*{4.5in} 
\vskip .2in
Figure 9e
\end{figure}

\begin{figure}
\plotone{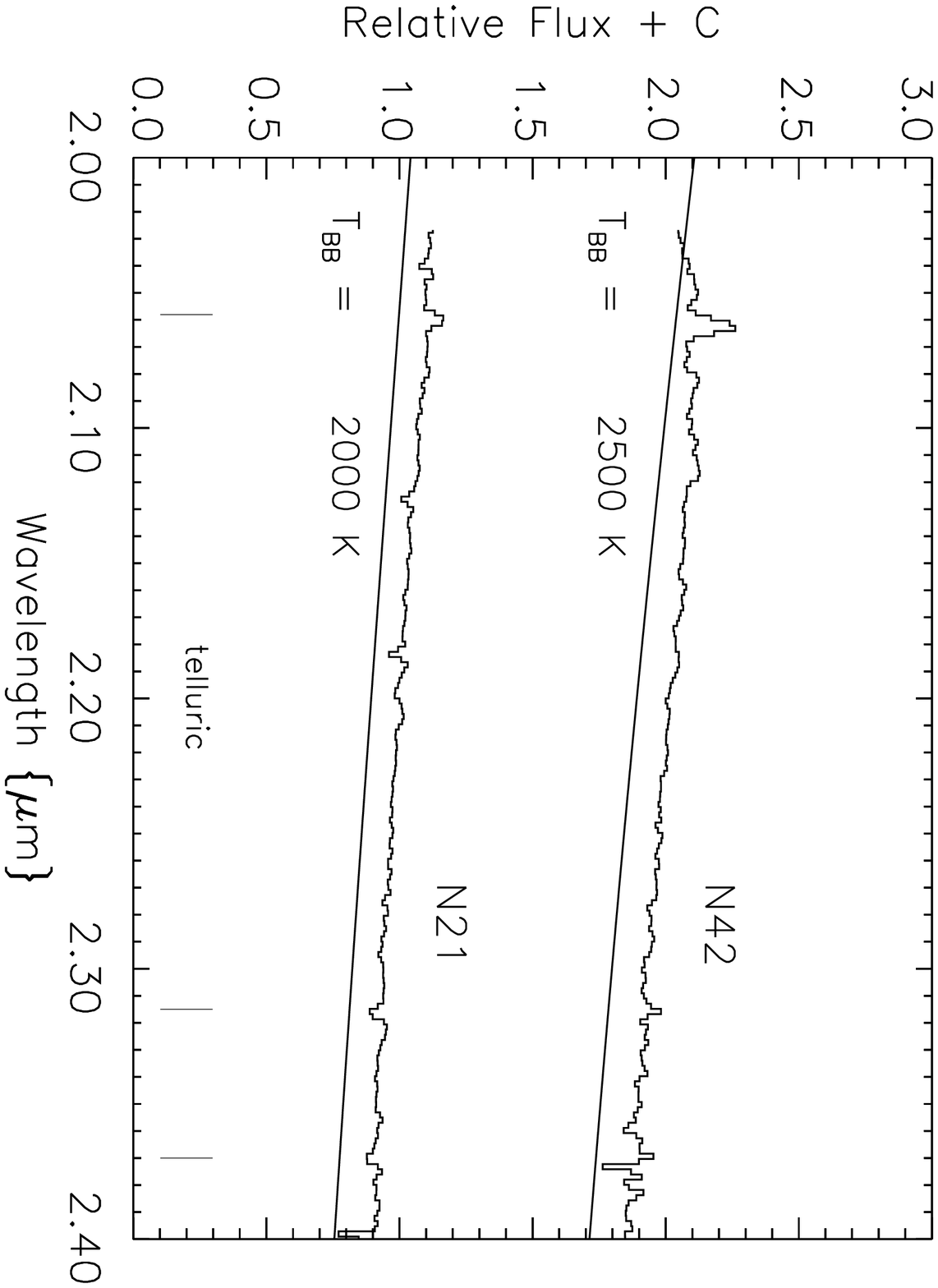}
\hspace*{4.5in} 
\vskip .2in
Figure 10
\end{figure}

\begin{figure}
\plotone{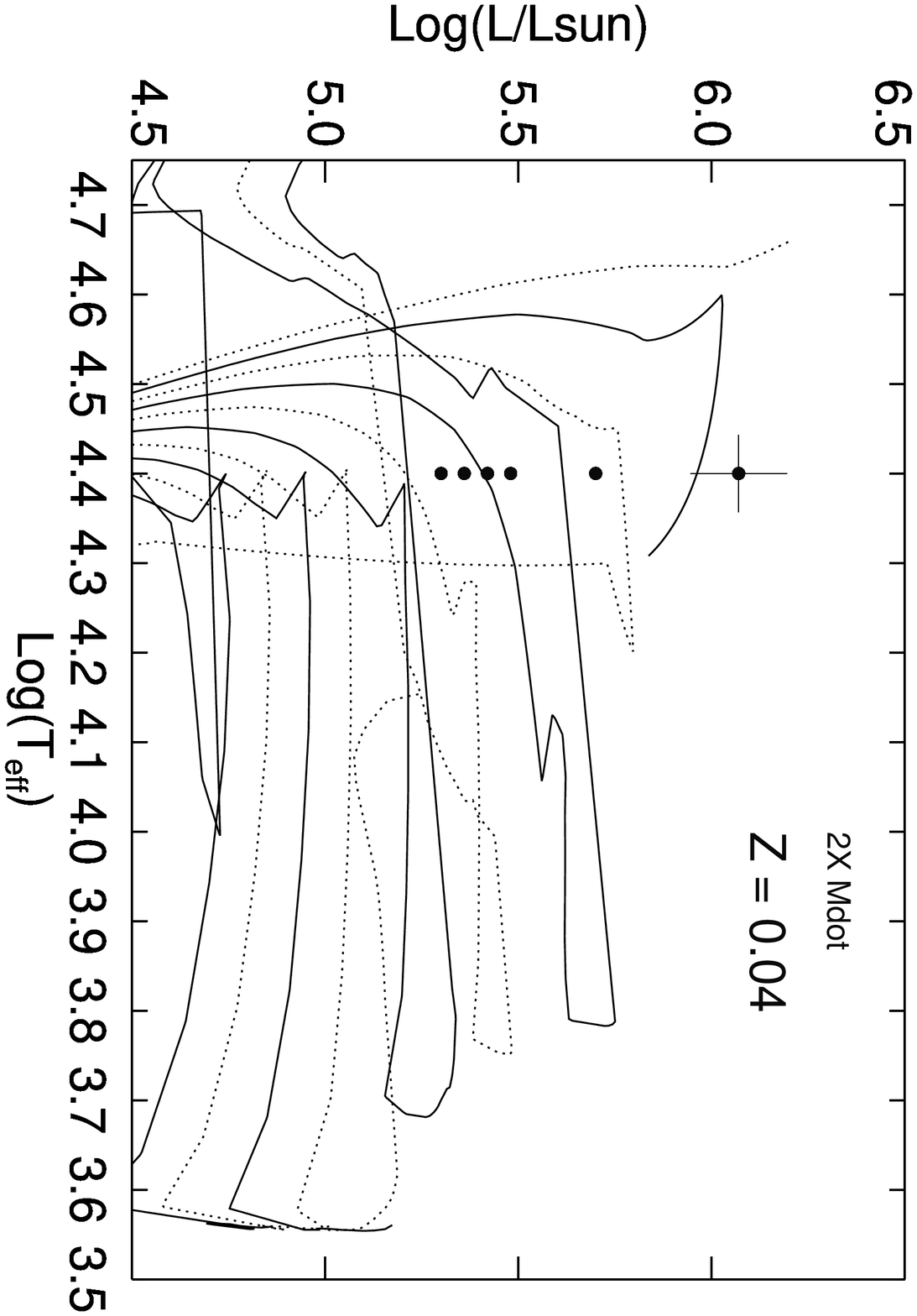}
\hspace*{4.5in} 
\vskip .2in
Figure 11
\end{figure}

\end{document}